\documentclass{article}

\usepackage{amsfonts}
\usepackage{array}
\usepackage{textcomp}
\usepackage{stfloats}
\usepackage{url}
\usepackage{verbatim}
\usepackage{graphicx}
\usepackage{subfigure}
\usepackage{balance}
\usepackage{bibentry}
\usepackage{hyperref}
\usepackage{algorithm}
\usepackage[dvipsnames, svgnames, x11names]{xcolor}
\usepackage{caption}
\usepackage{multirow}

\usepackage[algo2e,lined,boxed,commentsnumbered,ruled]{algorithm2e} 

\usepackage{lineno}

\usepackage{PRIMEarxiv}

\usepackage[utf8]{inputenc} 
\usepackage[T1]{fontenc}    
\usepackage{hyperref}       
\usepackage{url}            
\usepackage{booktabs}       
\usepackage{amsfonts}       
\usepackage{nicefrac}       
\usepackage{microtype}      
\usepackage{lipsum}
\usepackage{fancyhdr}       
\usepackage{graphicx}       
\graphicspath{{media/}}     

\title{FL-PLAS: Federated Learning with Partial Layer Aggregation for Backdoor Defense Against High-Ratio Malicious Clients
}

\author{
  Jianyi Zhang, Ziyin Zhou, Yilong Li, Qichao Jin\\
  Beijing Electronic Science and Technology Institute \\
  Beijing, China\\
  \texttt{zjy@besti.edu.cn} \\
   \And
  Xiali Hei \\
  University of Louisiana at Lafayette \\
  Lafayette, Louisiana, USA\\
  \texttt{email@email} \\
}

\begin{document}
\maketitle

\begin{abstract}
  Federated learning (FL) is gaining increasing attention as an emerging collaborative machine learning approach, particularly in the context of large-scale computing and data systems. However, the fundamental algorithm of FL, Federated Averaging (FedAvg), is susceptible to backdoor attacks. Although researchers have proposed numerous defense algorithms, two significant challenges remain. The attack is becoming more stealthy and harder to detect, and current defense methods are unable to handle 50\% or more malicious users or assume an auxiliary server dataset.

  To address these challenges, we propose a novel defense algorithm, FL-PLAS, \textbf{F}ederated \textbf{L}earning based on \textbf{P}artial\textbf{ L}ayer \textbf{A}ggregation \textbf{S}trategy. In particular, we divide the local model into a feature extractor and a classifier. In each iteration, the clients only upload the parameters of a feature extractor after local training. The server then aggregates these local parameters and returns the results to the clients.
  Each client retains its own classifier layer, ensuring that the backdoor labels do not impact other clients. We assess the effectiveness of FL-PLAS against state-of-the-art (SOTA) backdoor attacks on three image datasets and compare our approach to six defense strategies. The results of the experiment demonstrate that our methods can effectively protect local models from backdoor attacks. Without requiring any auxiliary dataset for the server, our method achieves a high main-task accuracy with a lower backdoor accuracy even under the condition of 90\% malicious users with the attacks of trigger, semantic and edge-case.
\end{abstract}


\section{Introduction}
Federated learning (FL) is a machine learning technique that has garnered significant attention due to its ability to protect data privacy in large-scale distributed systems. FL enables collaborative model training without the need to share sensitive data between multiple parties or a central server, making it particularly suitable for cloud and edge computing environments. In these systems, where data is distributed across multiple entities, ensuring privacy and security is critical. The technique allows data to remain secure on each participant’s device or server, avoiding the risks associated with centralized data storage \cite{konevcny2016federated,mcmahan2017communication}. Users train the model locally on their own data, upload updates to a central server, which then aggregates the models into a global model. This iterative process continues until the global model reaches a satisfactory level of accuracy. 

Privacy concerns have spurred the adoption of federated learning for collaborative training of shared deep learning models \cite{mcmahan2017communication}. However, this approach is vulnerable to backdoor attacks \cite{bouacida2021vulnerabilities,ozdayi2021defending}. These manipulated models function normally on clean data but maliciously misclassify backdoor samples \cite{chen2017targeted}. In this paper, we focus on trigger \cite{gu2017badnets}, semantic (like label-flipping) 
 \cite{bagdasaryan2020backdoor,tolpegin2020data,jebreel2024lfighter}, and edge-case attack \cite{wang2020attack} as they are challenging problems in FL. Other attacks like
directly modifies local weights and optimize the trigger pattern \cite{fang2023vulnerability}, or poisoning backdoor-critical layers \cite{zhuang2024backdoor} also achieved notable effects under specific conditions \cite{lyu2020threats}.

Existing defenses strategy are divided into two main types. One type aims to restrict model updates within controlled bounds through regularization or processing anomalous data. RSA \cite{li2019rsa} and NDC \cite{sun2019can} are examples of such defenses, which can weaken the impact of malicious updates on the model. However, this approach involves trade-offs, necessitating model normalization and noise introduction, which can affect accuracy, attack resilience, and main-task accuracy after being maliciously attacked. The other type relies on the classification of malicious models \cite{li2020learning} to detect and exclude the malicious local update, such as FLTrust \cite{cao2020fltrust}, FLAME \cite{nguyen2021flame}, and Krum \cite{blanchard2017machine}.

However, the effectiveness of these methods is based on the assumption that the majority of users are honest, which means that they may not work well when there is a large percentage (e.g., over 50\%) of malicious users.
Additionally, collecting some user data as an auxiliary dataset for the server, such as in the case of FLTrust, conflicts to some extent with the privacy protection of federated learning.

In this paper, we propose FL-PLAS (\textbf{F}ederated \textbf{L}earning based on \textbf{P}artial \textbf{L}ayer \textbf{A}ggregation \textbf{S}trategy), a backdoor defense algorithm. Our approach involves preserving the local classifier in the client, preventing contamination of benign users' local models by global backdoor neurons. This is achieved by dividing the local model into two parts: the feature extractor and the classifier. During each iteration, clients upload only the feature extractor's parameters after local training. The server then aggregates these parameters and shares the results with clients.  Each client keeps its own classifier layer to prevent malicious users' backdoor data from affecting benign users' models.


We experimentally evaluate the effectiveness of our FL-PLAS framework against trigger attacks \cite{gu2017badnets}, semantic attacks \cite{bagdasaryan2020backdoor}, and edge-case attacks \cite{wang2020attack}. Our results demonstrate that FL-PLAS can successfully defend against all of these attacks without compromising the privacy of user data.

Compared to the five defense methods (RSA, NDC, FLTrust, FLAME, and Krum), we find that most existing methods fail to effectively defend against backdoor attacks when the proportion of malicious users exceeds 50\%, except FLTrust and FL-PLAS. However, FLTrust requires the collection of user data, which conflicts with the privacy-preserving nature of federated learning.Specifically, our contributions are:

\begin{itemize}

    \item We conducted an in-depth analysis of the role different neural network layers play in defending against backdoor attacks. Using both simple and complex datasets, classification models, and extensive experiments with seven defense models and three attack models, we laid the foundation for understanding how specific layers contribute to backdoor defense.

    \item We implemented a backdoor defense scheme using the partial layer aggregation strategy, specifically targeting scenarios with a high proportion of malicious clients. Our approach effectively addresses the limitations of current methods, particularly in environments with a significant number of malicious clients and scenarios where the server cannot retain user data.

    \item Through rigorous experiments, we demonstrated that our method maintains superior performance even when the proportion of malicious clients exceeds 90\%. This highlights the robustness and efficacy of our approach in highly adversarial federated learning environments.

\end{itemize} 

\section{Background}
\label{background}
\subsection{Federated Learning (FL)}

\begin{figure}[ht]
\begin{center}
   \includegraphics[width=0.7\linewidth]{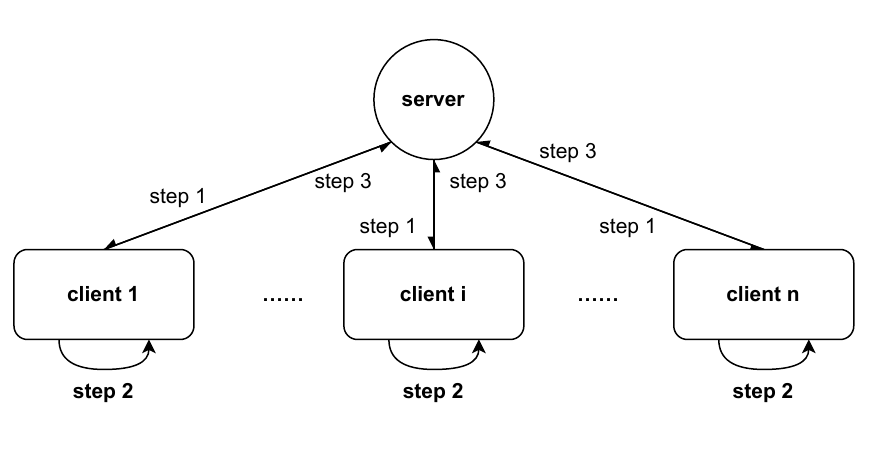}
\end{center}
   \caption{Illustration of the three steps in one iteration of FL. There are $N$ clients and a server. Each client has different classes and sizes of data, representing the heterogeneous distribution of client data.} 
\label{fig:FL}
\end{figure}

Federated learning is a type of distributed learning. As Figure \ref{fig:FL} shows, FL consists of $N$ users and one server. The user is responsible for training the model and passing the trained model to the server. And the server can aggregate the user's model to generate a global model. 


We roughly divide FL into three steps in one iteration (illustrated in Figure \ref{fig:FL}):
\begin{itemize}
    \item \textbf{Step 1}: The server sends the global model to the client, which gets the global model and starts training.
    
    \item \textbf{Step 2}: The client uploads the trained local model to the model aggregator, which is then aggregated by the server.
    \item \textbf{Step 3}: After the server-side aggregation is completed, the server sends the aggregated global model to the client.
\end{itemize}

\subsection{Backdoor Attacks in Federated Learning}
Generally, obtaining high accuracy on the testing dataset is the aim of the model design. Backdoor attacks do not destroy the accuracy of the global model, but induce it to make attacker-chosen mistakes on backdoor tasks \cite{chen2017targeted,dumford2020backdooring}. Backdoor attacks are applied in a variety of contexts, including crowd-sourcing systems \cite{fang2021data}, recommended systems \cite{fang2018poisoning,yang2017fake}, spam filters \cite{nelson2008exploiting}, etc. In the paradigm of FL, the server has no power to inspect the cleanliness of clients' data. Thus, a malicious client might poison his local data and train a malicious local model update based on it \cite{kairouz2019advances}. When the server frequently receives malicious local model updates from such clients, the global model obtained by aggregating the clients' local model updates will be compromised by the backdoor task. As a result, backdoor attacks pose a significant threat to the FL. Since implanting a specific backdoor into the model is much more complex and challenging, the current backdoor attacks commonly use data poisoning.

\textbf{Data poisoning \cite{li2020backdoor}}: The attacker can create backdoor data by adding triggers specified by the attacker (e.g., plus sign, etc.) and modifying the label of the data. This kind of data containing backdoor information is trained together with ordinary benign data. During the model's training process, the backdoor information from the local dataset gradually spreads to the global model, eventually leading to the poisoning of the global model.

Federated learning is more vulnerable to data poisoning attacks due to the larger number of users and the loss of supervision of user data due to the privacy-preserving nature of federated learning. The common data poisoning attacks are the trigger backdoor, the semantic backdoor, and the edge-case backdoor.


\textbf{Trigger backdoor \cite{gu2017badnets}}: The trigger backdoor attack is a prevalent type of backdoor attack that the attacker adds a specific mark or sign to an image and modifies its label. Consequently, the global model will misclassify other images that contain this specific mark. Typically, the triggers used in such backdoor samples are easily visible.

\textbf{Semantic backdoor \cite{bagdasaryan2020backdoor}}:
Semantic backdoor attacks are different from trigger backdoor attacks because they do not require any modifications to the input data. Instead, they exploit the semantic information already present in the model by assigning attacker-specified labels to data samples with specific features. For example, an attacker could assign the label ``cat'' to images of dogs with pointy ears, causing the model to misclassify images of other dogs with pointy ears as cats, or just poison the training data by flipping the labels from ``dog" to ``cat". This type of attack can cause the model to overlearn the features specified by the attacker, leading to inaccurate predictions for images that contain the same features.

\textbf{Edge-case backdoor \cite{wang2020attack}}: In the edge-case backdoor attack, the adversary utilizes extremely sparse data in the dataset and modifies its labels, causing the final model to misclassify such data that is hard to find in the training or test set. The attacker mainly exploits the robustness vulnerability in federated learning by modifying the labels of the sparse data in the user dataset. Since the data used in the edge case does not affect most of the data, this backdoor attack is more easily overlooked.

\subsection{Backdoor Defense Strategy}
\label{BD strategy}
There have been several proposed backdoor defense strategies that are effective against common backdoor attacks. The authors of papers \cite{he2020byzantine, hsu2019measuring} discuss why federated learning is vulnerable to backdoor attacks. 


\textbf{RSA \cite{li2019rsa}}:
For malicious updates of the model, we can regularize and control the updates mainly by using the model's update direction to control each model update within a certain range. Specifically, 
\begin{equation} \label{eq2} 
    G_{t} = \sum{\beta_{i}}sign(M_{i}-G_{0})
\end{equation}
where $M_{i}$ is the local model of client $i$, $G_{t-1}$ is the global model in last iteration, $\beta_{i}$ is a parameter mainly based on learning rate for $client_{i}$, and $G_{t}$ is the global model in this iteration.

\textbf{NDC \cite{sun2019can}}: Since backdoor updates may result in larger changes, model updates can be limited in scope by setting a threshold of $M$. The NDC (Norm Difference Clipping) method employs this strategy to limit the impact of backdoor attacks. The final global model update is calculated as:
\begin{equation} \label{eq3}
    u_{i}=\sum{\frac{u_{i-1}}{max(1,\frac{||u_{i-1}||}{TS})}}
\end{equation}
where $u_{i}$ is the model update after $i$ iterations, $||u_{i}||$ is the $l_2$-norm of the model update after $i$ iterations, and $TS$ is the threshold set by server.


\textbf{FLTrust \cite{cao2020fltrust}}: For FLTrust, the server needs to collect a small clean training dataset. During the training process, the server calculates the cosine similarity between the updates of the clean model and the updates of the user model.

\textbf{FLAME \cite{nguyen2021flame}}: FLAME utilizes a clustering approach to identify and remove adversarial model updates. It performs classification analysis and selects the models in the larger class as benign models for aggregation. FLAME calculates the Euclidean distance between the user and global models, takes the median as the benchmark, and calculates the weighted average of the model updates. To attenuate the backdoor, FLAME adds a certain amount of Gaussian noise.
Model weights are defined as:

\begin{equation} \label{eq4}
    e_{i}=\min(1,\frac{S_{median}}{S_i})
\end{equation}

where $e_{i}$ is the weight of clients, $S_{median}$ is the median Euclidean distance between the user model and the global model, and $S_{i}$ is Euclidean distance between the user model and the global model of client $i$.

\textbf{Krum \cite{blanchard2017machine}}:
Krum assumes that the server knows the number of pairs of malicious users and then calculates the model similarity among the models uploaded by users. Krum first finds the geometric center of the user models and then selects the models most similar to other user models as the final global aggregation models. Specifically,
\begin{equation} \label{eq5} 
    G = \mathop{\arg\min}||M_i-M_j|| 
\end{equation}

where $G$ is the global model, $M_i$ is the model of client $i$, $||M_i-M_j||$ is the Euclidean distance between the model $i$ and the model $j$. The key differences between our method and other defense strategies can be seen in Table \ref{comparie}.

%
\begin{table}[]
\footnotesize‌
\centering
\caption{Comparison of our method with other approaches}
\begin{tabular}{ccl}
\hline
                                  & Methods    & Defense strategy                                                                                                        \\ \hline
Averaging                         & FedAvg     & None                                                                                                                    \\ \cline{2-3} 
\multirow{3}{*}{Similarity-based} & Krum       & \begin{tabular}[c]{@{}l@{}}Calculates the Euclidean distance\\ Selects one local model\\ as the global model\end{tabular} \\ \cline{3-3} 
                                  & FLTrust    & \begin{tabular}[c]{@{}l@{}}Auxiliary dataset as reference\\ Cosine similarity with ReLU\\ to filter out\end{tabular}      \\ \cline{3-3} 
                                  & FLAME      & \begin{tabular}[c]{@{}l@{}}Calculates the cosine similarity\\ Applied a clustering\end{tabular}                         \\ \cline{2-3} 
Modify updates                    & RSA        & \begin{tabular}[c]{@{}l@{}}Calculates local update direction\\ Regularize and control the updates\end{tabular}          \\ \cline{2-3} 
Discard updates                   & NDC        & \begin{tabular}[c]{@{}l@{}}Sets an upper bound\\ Discards the larger model before\\ aggregation\end{tabular}              \\ \cline{2-3} 
Layer                             & Our Method & Partial model aggregation                                                                                               \\ \hline
\end{tabular}
\label{comparie}
\end{table}

\section{Related Work}
\label{related}

DNNs are vulnerable to both data and model attacks, including backdoor \cite{gu2017badnets,bagdasaryan2020backdoor}, evasion \cite{goodfellow2014explaining,papernot2016limitations},  fault injection \cite{hong2019terminal}, etc. In the backdoor attack, hidden triggers cause DNNs to make false predictions with attacker-specific data while behaving normally with benign data \cite{yang2017fake,rubinstein2009antidote,shafahi2018poison,suciu2018does,wang2019attacking}. Popular attack methods involve poisoning data or directly embedding triggers into models. A typical poisoning data attack involves inserting malicious samples or modifying training data to influence the behavior of the model \cite{chen2017targeted,gu2017badnets,fang2021data,fang2018poisoning, yang2017fake,nelson2008exploiting,rubinstein2009antidote,shafahi2018poison,suciu2018does,wang2019attacking,biggio2012poisoning,jagielski2018manipulating,li2016data,munoz2017towards,xiao2015feature}. Federated learning, due to its structure, is inherently more susceptible to data poisoning attacks \cite{bagdasaryan2020backdoor,wang2020attack,li2019rsa,he2020byzantine,fang2020local,baruch2019little,xie2020fall,bhagoji2019analyzing}. From the perspective of the purpose of the attack, these attacks can be divided into untargeted attacks \cite{li2019rsa,he2020byzantine,fang2020local,baruch2019little,xie2020fall} and targeted attacks \cite{bagdasaryan2020backdoor,wang2020attack,bhagoji2019analyzing,xie2019dba}. Untargeted attacks aim to deteriorate the global model. And targeted attacks aim to induce the global model to make some attacker-chosen mistakes in certain inputs, without deteriorating the global model. In terms of attack types, these attacks in federated learning are categorized as trigger, semantic, and edge-case attacks.

In order to cope with challenges of backdoor attacks, defense methods are essential. In backdoor defense, the goal of the defender is to minimize the impact caused by the backdoor. Defenses can be categorized by their mechanisms: trigger-backdoor mismatch \cite{liu2017neural,doan2020februus}, backdoor elimination \cite{zhao2020bridging,liu2018fine,tran2018spectral,chen2018detecting,tang2021demon,andreina2021baffle}, and trigger elimination \cite{gao2019strip,subedar2019deep,jin2020unified}. In federated learning, backdoor defenses divide into two types: mitigating the impact of malicious models on the global model (limitation) \cite{li2019rsa,sun2019can}, and detecting malicious models \cite{blanchard2017machine}. Limitation-based defenses constrain user-uploaded model updates' norms, minimizing malicious models' global impact. For the detection of malicious models \cite{blanchard2017machine}, servers identify and reject malicious models. Recent research has demonstrated inverting local model updates to exclude malicious updates from aggregation \cite{zhao2022fedinv}, and a reverse engineering-based trigger defense can provide a sufficient condition on the quality of trigger recovery \cite{zhang2023flip}.

However, it cannot be neglected that the effectiveness of these methods relies heavily on the assumption that the majority of users behave honestly. Assuming a condition that the number of dishonest users (malicious users) exceed over 50\%. Under such circumstance, these methods did not worked well at all. Moreover, some methods collect a certain amount of local data from users as the root dataset \cite{cao2020fltrust} or reference model \cite{raza2022using}. Since collecting these data might lead to the leakage of sensitive user information, the requirements are too strict for federated learning and may not be applicable to many scenarios. The way to solve these problems is very important.

Existing solutions, such as classification and clustering-based methods, have significant limitations. Classification-based methods struggle when malicious users are in the majority, and clustering methods become ineffective as they rely on the assumption of fewer malicious clients. Moreover, solutions that depend on the server retaining sample data are impractical due to strict data privacy requirements and the inherently distributed nature of data.

Regarding pFL-related research \cite{collins2021exploiting, chen2024efficient, pillutla2022federated, arivazhagan2019federated, t2020personalized, tan2023pfedsim}, while Gao \cite{gao2020end} and Qin et al. \cite{qinpfl2023} demonstrated that partial model aggregation could effectively defend against backdoor attacks, their study merely provided a simple evaluation of various pFL algorithms initially designed to address data heterogeneity. They did not conduct detailed research on the effectiveness of pFL against different proportions of malicious clients, different model parameters, nor did they deeply explain the reasons for its effectiveness.
\section{Problem Setup}
 \label{problem}
\textbf{Threat model:} Similar to \cite{fang2020local,cao2020fltrust,huang2020one}, we make the following assumptions about the attacker:
(i) Attacker controls one or multiple users, replacing original data with backdoor data via user data modification; (ii) Attacker obtains complete info of controlled user - user data, loss function, learning rate; (iii) Malicious users select data for training, manipulate local model updates at will - modify local model training's learning rate, scale model update; (iv) Malicious users can attack any typical deep neural network (DNN); (v) Malicious users do not need to be omniscient and collude with each other.

\textbf{Defense goals:}
Similar to \cite{cao2020fltrust}, we evaluate our method via \textbf{fidelity}, \textbf{robustness}, and \textbf{efficiency}. For fidelity, we target maximal benign update retention during non-attack states. For robustness, our method's efficacy is expected across scenarios with diverse malicious user ratios, backdoor attack types, etc. Efficiency-wise, we refrain from imposing excessive computation and memory demands on user devices beyond FedAvg.

\textbf{Defender’s knowledge and capability:} 
The backdoor defense scheme herein is essentially an aggregation rule. Since this aggregation rule mainly runs on the server side, the defender acquires server-side information, including the global model, user-uploaded model parameters, and user count. However, the server side remains uninformed regarding user data and attacker details, such as malicious user ratio or attack type. Additionally, the server cannot collect user data as in FLTrust. Such data collection could risk exposing sensitive user information. Consequently, the proposed algorithm's operational context is more discreet and pragmatic compared to FLTrust.

\textbf{Evaluation Metrics}\label{EM}: For defense methods, we test the global model using a clean test set and a backdoor test set, respectively. To obtain the main-task accuracy (MA) and backdoor accuracy (BA), the backdoor test set is composed by adding a backdoor to the clean test set. We assume that there are $BU$ benign users among all users, $M_{i}$ are the samples of the $i-th$ client whose labels are correctly predicted on the clean test set, and $|M_{i}|$ is the size of $M_{i}$. $B_{i}$ are the samples of the $i-th$ client whose labels are classified as the attacker-chosen class (backdoor attack successfully attacked), and $|B_{i}|$ is the size of $B_{i}$. $|M|$ and $|B|$ are the size of clean test set and backdoor test set. 

Here we define MA as the proportion of data the model predicts correctly on a clean test set. That is, $MA=\frac{1}{BU}\sum{\frac{|M_{i}|}{|M|}}$. So, the higher the MA is, the more correctly the model predicts. BA is the proportion of data in the backdoor task that the global model classifies as the attacker-chosen class. That is $BA=\frac{1}{BU}\sum{\frac{|B_{i}|}{|B|}}$. Therefore, the higher the BA is, the more successful the attack is, which indicates a weaker capability of the defense solution.. For FL-PLAS, we test all local models and average their MA and BA to be the MA and BA of FL-PLAS.

\section{FL-PLAS Overview and Design}
\label{overview}
\subsection{High-level Idea}
\textbf{Motivation:} In federated learning, preventing backdoor data from poisoning benign client models is crucial. Common backdoor defense algorithms typically rely on the classification or comparison of user models with benign server models. However, when the number of malicious users exceeds 50\%, classification-based methods become challenging to handle. In such settings, traditional FL defense mechanisms often fail. Some methods rely on clustering to identify malicious clients, which becomes ineffective when the majority of clients are malicious. Others require the server to maintain a portion of the sample data, which is impractical in many computing and data systems where data privacy and distribution uniformity cannot be guaranteed.

There is a clear gap in developing robust FL defense mechanisms that can effectively operate in environments with a high proportion of malicious clients and where data privacy is paramount. Current methods do not adequately address the challenges posed by highly adversarial conditions and stringent privacy constraints. Our previous research observes that some certain layers in neural networks exhibit distinguishable patterns between malicious and benign updates \cite{xmam}. Hence, we hypothesize that processing certain layers separately may break the connection between backdoor data and its corresponding labels. We try to isolate the influence of backdoor clients during the training of federated learning models, ensuring that their impact remains confined to their own clients and does not affect benign clients.


\textbf{Key observation and idea:} 
In order to test our hypothesis, we conducted a simple experiment comparing the main-task accuracy and backdoor accuracy of four types of neural network models. We first trained a clean model and a backdoor model on the MNIST dataset \cite{lecun1998gradient} using the \texttt{Lenet} \cite{lecun1989handwritten} architecture.
Then we separated these two models into their corresponding feature extractors (FE) and classifiers, and assembled them into four new models: clean FE with clean classifier, clean FE with backdoor classifier, backdoor FE with clean classifier, and backdoor FE with backdoor classifier.

As shown in Table \ref{tab:intuition}, the backdoor accuracy is dependent on whether the classifier is poisoned or not, regardless of whether the feature extractor is poisoned or not. The same observations are also demonstrated on the CIFAR-100 dataset \cite{krizhevsky2009learning} with \texttt{ResNet-18} \cite{he2016deep}. Hence, the key idea of our algorithm is that the server only aggregates part of the model uploaded by the user, and the user keeps their classifiers locally.

\begin{table}[htbp]
\footnotesize‌
\caption{The main-task accuracy (MA) and backdoor accuracy (BA) of neural networks with backdoor in different Feature Extractors (FE) and Classifiers.}
\small
\renewcommand\arraystretch{1.5}
\centering
\begin{tabular}{c|cc}
\hline
MA/BA               & Clean FE & Backdoor FE \\ \hline
Clean Classifier    & 0.99/0.1                & 0.99/0.1                     \\
Backdoor Classifier & 0.98/1                  & 0.98/1                       \\ \hline
\end{tabular}

\label{tab:intuition}
\end{table}


\subsection{FL-PLAS Design}
According to our interesting findings, we divide our model into two parts: the feature extractor and the classifier. In each iteration, the clients only upload the parameters of the feature extractor after local training. Then, the server aggregates these local parameters and returns the results to the clients. With the partial layer aggregation strategy, every client keeps its own classifier layer to isolate the malicious users' backdoor data from benign users.

\begin{figure}[t]
\begin{center}
   \includegraphics[width=0.6\columnwidth]{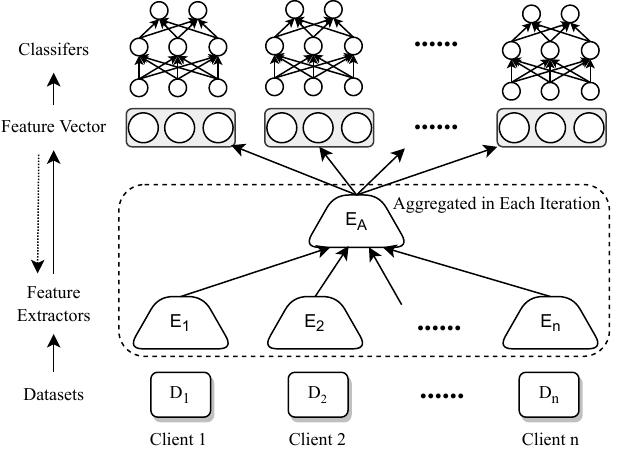}
\end{center}

\caption{Illustration of FL-PLAS workflow in round $t$.}
\label{fig:FL-PLAS}
\end{figure}


    
As the data distribution is unknown to us, we use empirical risk as our risk function. The risk function of client $i$ is :
\begin{equation} \label{eq7} 
    R_{exp}(f)=\frac{1}{S} \sum_{i=1}^{S}(L(y_i,f(x_i)))
\end{equation}
    where $S$ is the sample content of client $i$.

For example, if we choose classifier as our cutting layer, we can divide the model into feature extractor (FE) and classifier (CL). This is how model works.
\begin{equation} \label{eq8} 
    \hat{y_{i}}=FE(x_{i})
\end{equation}
\begin{equation}
        y_{i} =CL(\hat{y_{i}})
\end{equation}
where $x_{i}$ is the data of client $i$, $\hat{y_{i}}$ is the feature extraction result (or embedding), and $y_{i}$ is the prediction result.

We upload the feature extractor part of the model, which is aggregated by the aggregator, while keeping the classifier locally on the user side. We denote the number of clients as $r$.
\begin{equation} \label{eq9} 
    G = FedAvg(M_{1,FE},\cdot\cdot\cdot,M_{r,FE})
\end{equation}
\begin{equation}
    M_{i} = G +M_{i,CL}
\end{equation} 
Figure \ref{fig:FL-PLAS} and Algorithms \ref{alg:FL-PLAS} and \ref{alg:ClientUpdate} illustrate the working mechanism of FL-PLAS in round $t$.
 
\textbf{Step 1}: User $i$ uses their local dataset $D_{i}$ to train a local model $M_i$ and then uploads their model to the server.

\textbf{Step 2}: After receiving the models uploaded by the users, the server aggregates specific layers of the models (e.g., feature extractors) to obtain the global model $G$. The server then returns the aggregated model to the users.

\textbf{Step 3}: Users replace their partial model with the global model $G$, then use their local dataset $D_i$ and the new local model to calculate the feature vectors and losses. They then use these to optimize the entire model, including both the feature extractor and classifier.

\begin{algorithm}
\small
	\caption{FL-PLAS (aggregation rule)}
	\label{alg:FL-PLAS}
    \KwIn{The number of received clients $r$, number of cutting layers $l$, total iterations $T$, initial model $G^{0}$, size of local examples $n$}    
    \KwOut{The global model ${G}$.} 
	\BlankLine
	\For{$t$ = 1, 2, ···, $T$}
	{
	   \For{$i$ = 1, 2, ···, $r$}
	        {$M_{i}^{t}\gets ClientUpdate(G^{t-1},l) $}
	   \For {layer $p$ in $M_{i}$}
            {$if$ $p<l$ $then$ $G_{p}^t\gets \sum{\frac{n_{i}}{n}M_{i,p}^t}$        }     
	} 
	return $G$
\end{algorithm}

\begin{algorithm}
\small
	\caption{ClientUpdate}
	\label{alg:ClientUpdate}
    \KwIn{The local dataset of client i $D_{i}$, global model from server $G$, number of cutting layers $l$.}
    \KwOut{The local model ${M_i}^{t}$.} 
	\BlankLine
	   \For {layer $p$ in $M_{i}$}
            {$if$ $p<l$ $then$
            $M_{p}^t\gets G_{p}$}            
        \For{each batch $b\in D_{i}$}
            {
            $M_{i}\gets M_{i}-\eta \Delta l(b,M_{i})$, $\eta$ is the local learning rate, $\Delta l$ denotes the loss using data $b$ and model $M_{i}$
            }
	return ${M_i}^{t}$  
\end{algorithm}

%
The algorithm complexity of FL-PLAS covers two aspects, global model update and client update (Algorithm 2). According to Algorithm 2, the main complexity of client update is $ O(p+\frac{D_i}{b}) $, which relies on two iterations with $p$ times and $\frac{D_i}{b}$ times. Therefore the complexity of Algorithm 2 equals to Linear complexity $O(N)$. According to Algorithm 1, the main complexity of global model update is $ O(p*n) $, which covers $n$ models to aggregate with $p$ times iteration. Therefore, the complexity of FL-PLAS is $O(r*O(N)+p*n) $ and equals to $O(N^2)$.

\begin{table*}[h]
\caption{The default FL system parameter settings.} 
\vspace{-5mm}
\footnotesize‌
\centering
\setlength{\tabcolsep}{1mm}{
\begin{center}
\begin{tabular}{|c|c|c|c|c|c|}
\hline
Dataset and Partition &MNIST &\multicolumn{3}{c|}{CIFAR-10} &CIFAR-100  \\          \hline
 Total number of clients&\multicolumn{5}{c|}{100}\\ 
 \hline
 Client per round&	\multicolumn{5}{c|}{30}\\ 
 \hline
 Backdoor type& Trigger & Trigger& Semantic& Edge-case&Trigger\\ 
 \hline
 Learning rate&\multicolumn{1}{c|}{$6.7\times10^{-3}$}&\multicolumn{3}{c|}{$2.7\times10^{-3}$}&	$1.5\times10^{-5}$ \\ 
 \hline
 Local iterations & \multicolumn{5}{c|}{1} \\ 
 \hline
  Global training round &\multicolumn{5}{c|}{200}  \\ 
  \hline
  Batch size&\multicolumn{5}{c|}{32} \\ 
  \hline
  Combined learning rate&	\multicolumn{5}{c|}{learning rate $\times 0.998^t$} \\ \hline 
Optimizer&	\multicolumn{5}{c|}{SGD}\\ 
\hline
Momentum&	\multicolumn{5}{c|}{0.9}\\ 
\hline
Weight decay&	\multicolumn{5}{c|}{$10^{-4}$}\\ 
\hline
\end{tabular}
\label{tab:setting}
\end{center}
}
\end{table*}

\section{Evaluation}
\label{evaluation}

\subsection{Experimental Setup}
 
1) \textit{Datasets}: We use three image datasets to evaluate the effectiveness of FL-PLAS. The datasets are also divided according to the number of users (M). For the non-independent and identically distributed (non-i.i.d.) data, we divide the data according to the Dirichlet distribution, where the users get different data in terms of distribution and data volume. In addition, the parameter used for the Dirichlet distribution in this experiment is 0.2, \textit{i.e.}, the data distribution $\chi \sim Dir(0.2, M)$. 

\textbf{MNIST:} The MNIST dataset is widely used in computer vision tasks and consists of handwritten digits. It contains a training set of 60,000 samples and a testing set of 10,000 samples, which are normalized to 28$\times$28 pixels and centered at a fixed size.

\textbf{CIFAR-10:} The CIFAR-10 dataset is a color image classification dataset that consists of 50,000 training images and 10,000 test images. Similar to MNIST, the CIFAR-10 dataset classifies the images into 10 categories, which include airplanes, cell phones, and birds.

\textbf{CIFAR-100:} CIFAR-100 is very similar to CIFAR-10, but it contains 100 classes instead of 10. Each class in CIFAR-100 contains 500 training images and 100 test images.

For comparability, we used a consistent experimental setup. The MNIST dataset employed the trigger attack for the backdoor, employing a \texttt{Lenet} model. CIFAR-10 evaluated FL-PLAS against various attacks (trigger, semantic, edge-case) using a \texttt{MobileNet} model \cite{howard2017mobilenets}. CIFAR-100 focused on the trigger attack, employing a \texttt{ResNet-18} model.

The MNIST dataset tested FL-PLAS's backdoor defense on simple images and validated its performance on the \texttt{Lenet} model. CIFAR-10, coupled with \texttt{MobileNet}, assessed defense effectiveness on color images and complex models, demonstrating robustness against new attacks. CIFAR-100 showcased backdoor defense across datasets with up to 100 classes.

 2) \textit{Attack settings}: In our experiment, we evaluate the effectiveness of FL-PLAS against three types of backdoor attacks: the trigger attack, the semantic attack, and the edge-case attack. In each attack, the attacker applies a backdoor processing technique to a certain percentage $p$ of user data to inject backdoor patterns for subsequent attack processing.

\textbf{Trigger attack:} The trigger attack modifies user data by adding the same trigger logo to each sample (a 2$\times$2 white box at the top right for MNIST and a 5-pixel white plus sign at the top right for CIFAR-10/CIFAR-100) and changing the label to a specified attacker label (in our experiment, the attacker label is set to 0).


\textbf{Semantic attack:} As in \cite{bagdasaryan2020backdoor}, in the training and evaluation datasets, we select the ``green car" as the backdoor images. We modify the label of the ``green car" to be ``bird" for this attack.

\textbf{Edge-case attack:}
As in \cite{wang2020attack}, we use aircraft images oriented in the southwest direction to generate the backdoor data and the backdoor test set. The southwest-oriented aircraft represent an extremely small percentage of the global data. For the generated backdoor data, we modify its data label to category 9 for trucks. Additionally, if the backdoor is not triggered, the above backdoor data will be classified as category 1 for cars.

\begin{figure*}[ht]
\setlength{\tabcolsep}{0.8mm}{
\begin{center}
\subfigure[MNIST\label{fig:bamnist}]
{\includegraphics[width=0.32\linewidth]{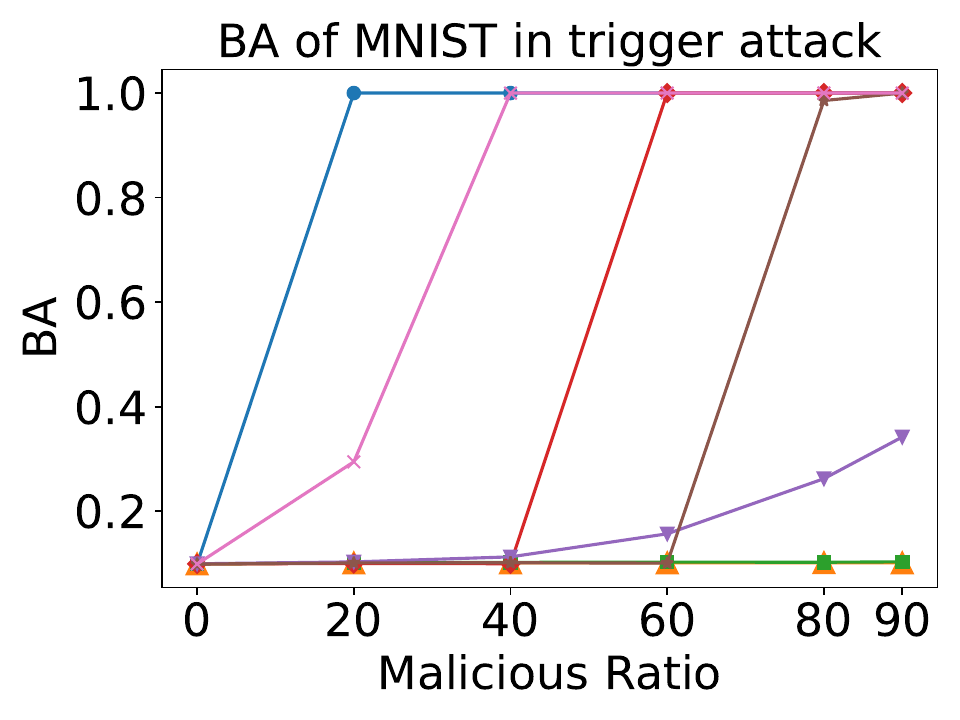}}
\subfigure[CIFAR-10\label{fig:baCIFAR10}]
{\includegraphics[width=0.32\linewidth]{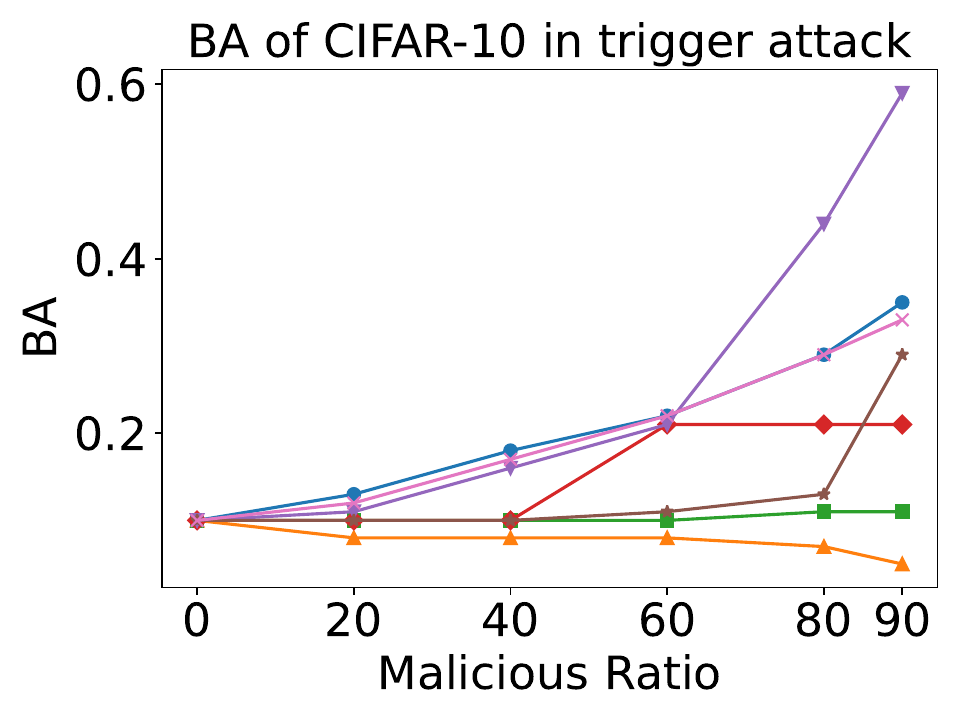}}
\subfigure[CIFAR-100\label{fig:baCIFAR100}]
{\includegraphics[width=0.32\linewidth]{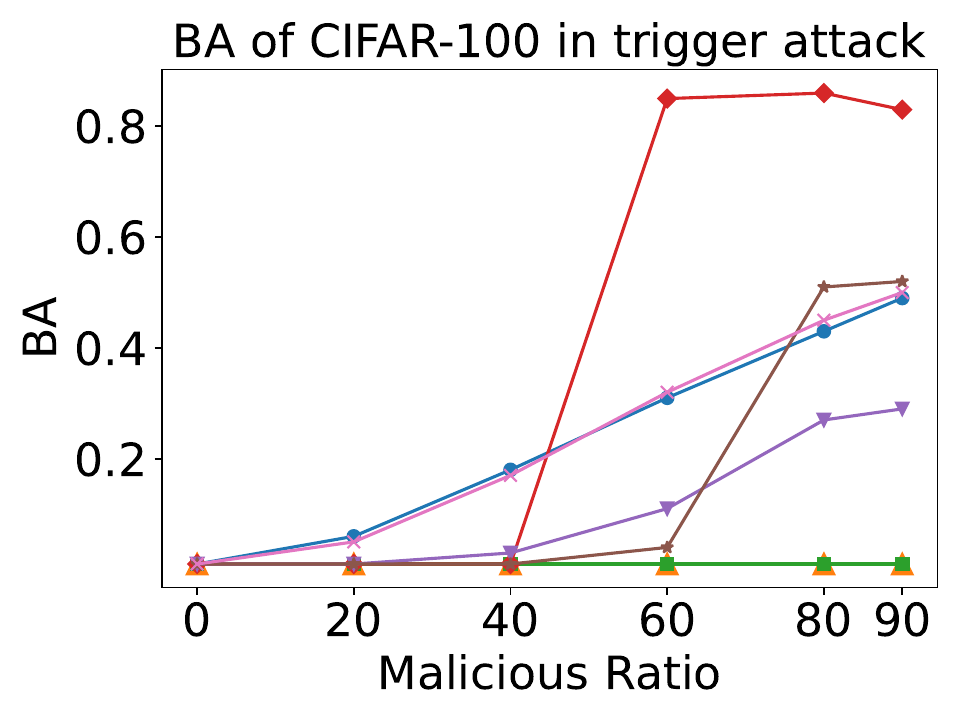}}
\end{center}}
\begin{center}
\vspace{-5mm}
\includegraphics[width=1\linewidth]{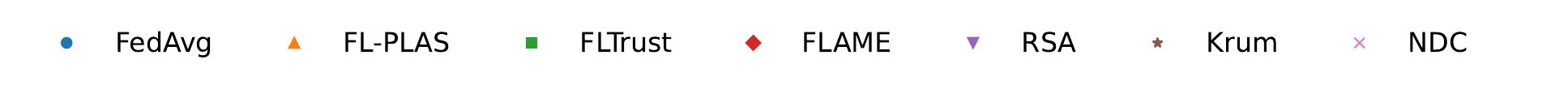}
\end{center}\vspace{-8mm}
\caption{BA of various datasets under the trigger attack. \ref{fig:bamnist} depicts how the BA of MNIST changes against different ratios of malicious clients; \ref{fig:baCIFAR10} depicts how the BA of CIFAR-10 changes; and \ref{fig:baCIFAR100} depicts how the BA of CIFAR-100 changes.}
\label{fig:baCIFAR10`}
\end{figure*}

3) \textit{System Settings}: In this experiment, we use the same experimental settings as in \cite{gu2017badnets}. We set the total number of users to 100 and the number of users selected each time to 30\%, where a constant number of malicious users appear in each training according to the proportion of malicious users. Moreover, the malicious users will backdoor 30\% of their data to generate backdoor data. The detailed model parameters are shown in Table \ref{tab:setting}.

4) \textit{Defenders' Settings}: 
For FLtrust, we set the size of the clean small training dataset (called ``root dataset") to be $100$ and the local iteration to be $1$ as \cite{cao2020fltrust}.
For FLAME, we set $\sigma$ (the noise level bound in FLAME) to 0.01 according to \cite{nguyen2021flame}. 

\subsection{Experimental Results}\label{sec:exp}

\subsubsection{Robustness}
Figure \ref{fig:bamnist} illustrates the defense effects against backdoor attacks on the non-i.i.d. MNIST dataset. Existing defense strategies show satisfactory results when the proportion of malicious users is under 40\%. On the contrary, FL-PLAS, FLTrust, and Krum outperform others when malicious users exceed 50\%. FLAME is a cluster-based method that performs the best when the number of malicious users is less than 40\%, and its performance gradually declines when the malicious users exceeds 40\%. The performance of RSA is decreasing as the number of malicious users is increasing. Krum is effective with 60\% malicious users due to pre-trained models aiding convergence and benign model selection. However, Krum becomes ineffective as malicious users reach 80\%.

Figure \ref{fig:baCIFAR10} illustrates the pronounced superiority of our approach in the CIFAR-10 dataset. All six strategies exhibit robust backdoor defense below 50\% malicious users proportion. However, as malicious users increasing, only FL-PLAS and FLTrust maintain better defense.

Figure \ref{fig:baCIFAR100} illustrates a more substantial disparity among defense methods in the non-i.i.d. CIFAR-100 dataset. Under 40\% malicious users, FL-PLAS, FLTrust, FLAME, and Krum exhibit stronger defense. Only FL-PLAS and FLTrust sustain effective defense against backdoor attacks when the malicious users exceed 40\%

\subsubsection{Two New Types of Attacks}

\begin{figure}[ht]
\begin{center}
\subfigure[semantic\label{fig:baseman}]{   \includegraphics[width=0.36\linewidth]{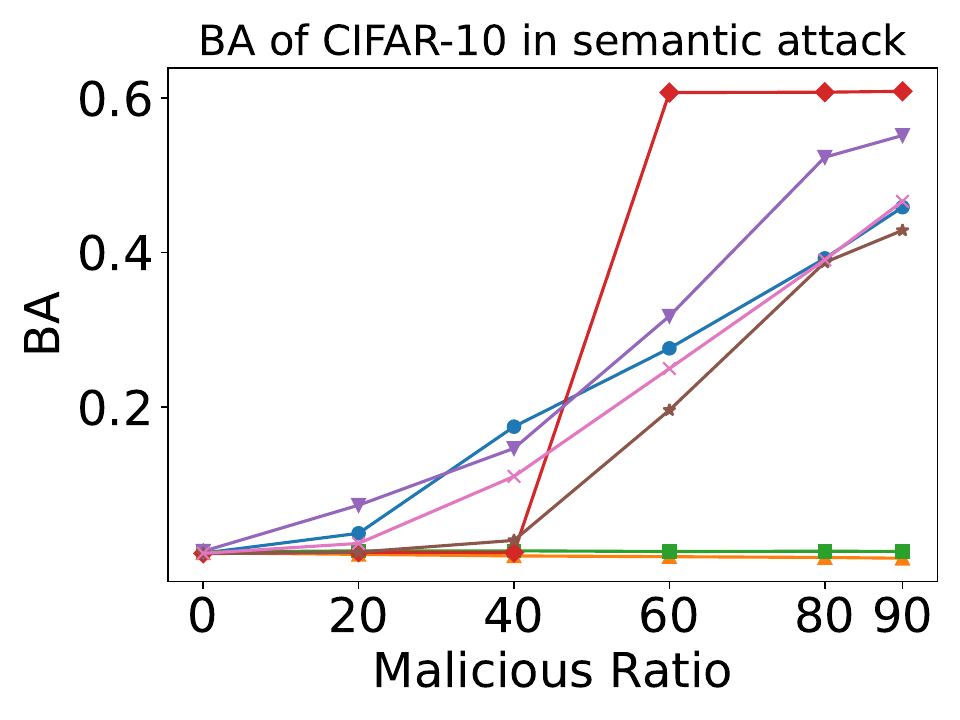}}
\hspace{20pt}
\subfigure[edge-case\label{fig:baedge}]{   \includegraphics[width=0.36\linewidth]{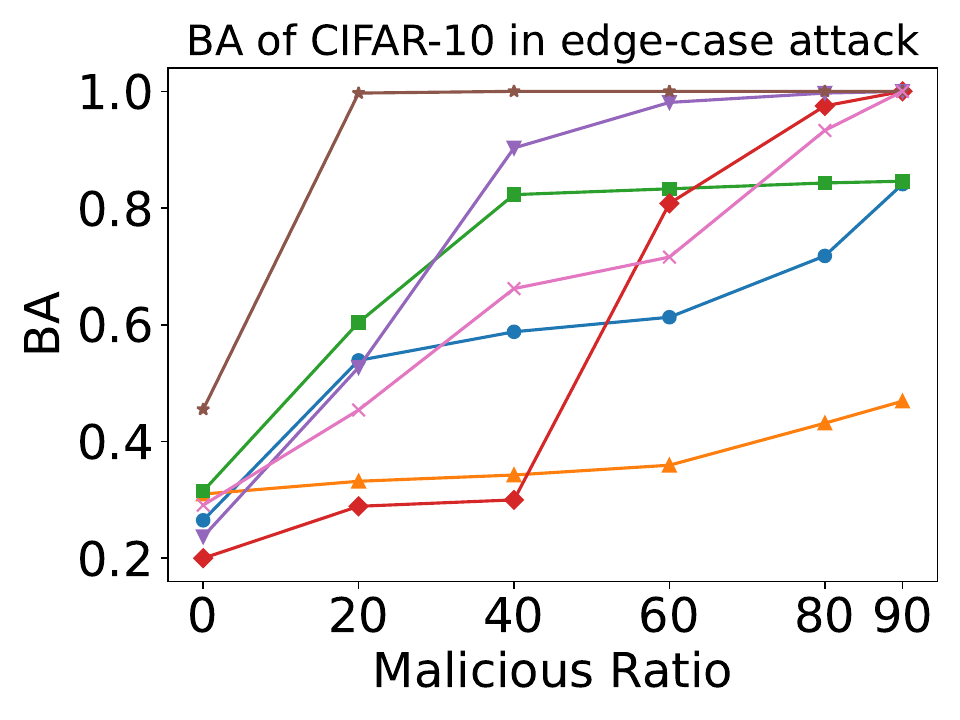}}
\begin{center}
\includegraphics[width=1\linewidth]{figure/legend_fig3.pdf}
\end{center}
\end{center}
\vspace{-8mm}
   \caption{BA of CIFAR-10 in two new types of attacks. \ref{fig:baseman} shows semantic attack, \ref{fig:baedge} shows edge-case attack.}
\label{fig:baedge0}
\end{figure}

Besides the basic trigger attacks, there are two new backdoor attacks emerging recently: semantic attacks and edge-case attacks. Taking the CIFAR-10 dataset as an example, the effectiveness of the above mentioned backdoor defense strategies for the new types of attacks is shown in Figure \ref{fig:baedge0}.

As it shown in Figure \ref{fig:baseman}, for the semantic attack on the CIFAR-10 dataset, FL-PLAS and FLTrust can still ensure the backdoor content remaining within a reasonable range even being attacked by 80\% of malicious users. FLAME and Krum show a strong ability to identify malicious models when the proportion of malicious users is less than 40\%. When the malicious users is higher than 40\%, the defense effect becomes weaker. Since the strategies of RSA and NDC fail to identify malicious models and benign models under semantic attacks, the global models of these methods gradually lose their resistance against backdoor attacks, which causes the value of BA in Figure \ref{fig:baseman} increasing gradually, and indicates an increase in the backdoor content in the model.

For edge-case attacks, as shown in Figure \ref{fig:baedge}, most of the backdoor defense methods are difficult to defend against edge-case attack when the proportion of malicious users is high. Only FL-PLAS performs good defense capability against edge-case backdoor. FLAME prevents the improvement of model backdoor task accuracy by aggregating benign users and adding noise when the proportion of malicious users is less than 40\%. When the proportion of malicious users is large, malicious users will also participate in aggregation, resulting in an increase in backdoor accuracy.

\begin{table}[htb]
\footnotesize‌
\caption{The BA when there are 90\% malicious users.}
\begin{center}
\vspace{-5mm}
\scalebox{1}{
\subtable[Trigger attack]{
\setlength{\tabcolsep}{0.7mm}{
\begin{tabular}{ccccccccc}
\hline
&FedAvg&FLTrust&FLAME&RSA&Krum&NDC&FL-PLAS\\
\hline
MNIST     & 1      & 0.10    & 1     & 0.34 & 1    & 1    & \textbf{0.10} \\
CIFAR-10  & 0.35   & 0.11    & 0.21  & 0.59 & 0.29 & 0.33 & \textbf{0.05} \\
CIFAR-100 & 0.49   & 0.01    & 0.83  & 0.29 & 0.52 & 0.50  & \textbf{0.01} 
\\
\hline
\end{tabular}}}}
\scalebox{1}{
\subtable[New types of attack]{
\setlength{\tabcolsep}{0.7mm}{
\begin{tabular}{cccccccc}
\hline
&FedAvg&FLTrust&FLAME&RSA&Krum&NDC&FL-PLAS\\
\hline
Semantic  & 0.46   & 0.01    & 0.61  & 0.55 & 0.43 & 0.47 & \textbf{0.004} \\
Edge-case & 0.84   & 0.85    & 1     & 1    & 1    & 1    & \textbf{0.47} 
\\
\hline
\end{tabular}}}}
\label{tab:0.9BA}
\end{center}
\end{table}

When the proportion of malicious users is up to 90\%, as shown in Table \ref{tab:0.9BA}, our method FL-PLAS can still be effective in reducing the backdoor in the model.Moreover, our method improves defense capability by over 44\% compared to other methods, reaching up to 99.34\% (FLAME).

\subsubsection{Main-task Accuracy}

An important evaluation factor for a defensive strategy is how well it performs on the main task while simultaneously restricting the backdoor.

\begin{figure*}[ht]
\setlength{\tabcolsep}{0.5mm}{
\begin{center}
\subfigcapskip=-3pt
\subfigure[MNIST\label{fig:mamnist}]{   \includegraphics[width=0.315\linewidth]{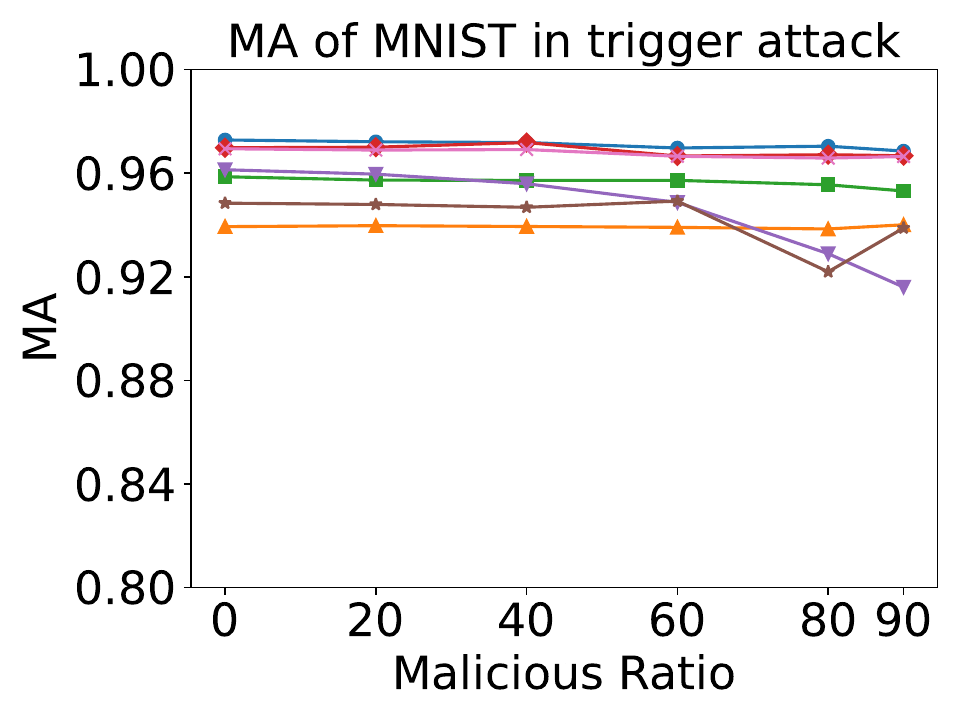}}
\subfigure[CIFAR-10\label{fig:maCIFAR10}]{   \includegraphics[width=0.315\linewidth]{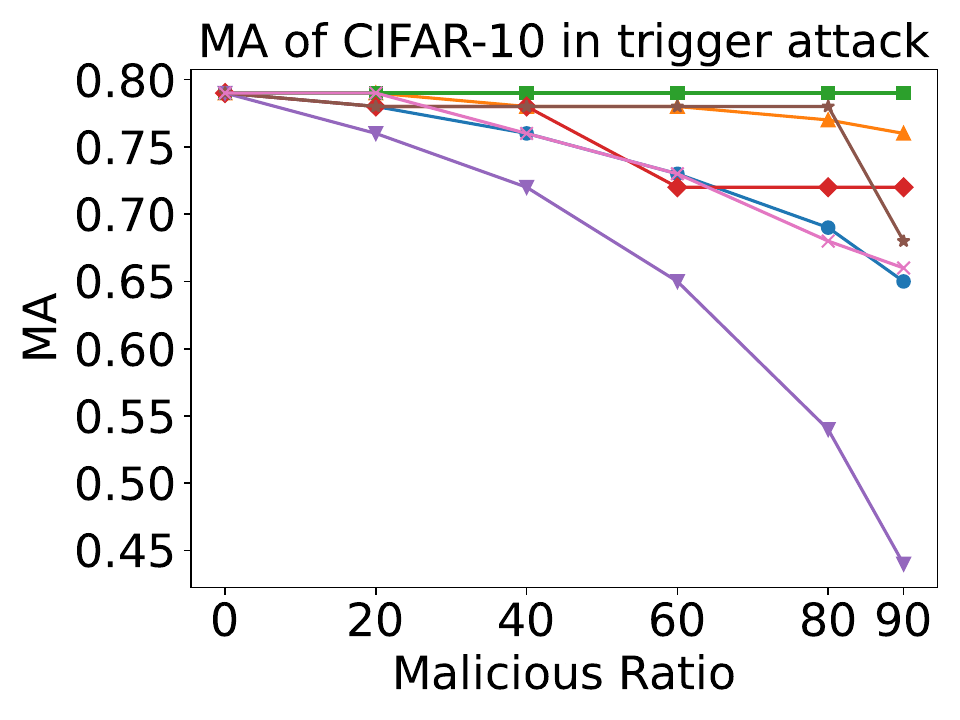}}
\subfigure[CIFAR-100\label{fig:maCIFAR100}]{   \includegraphics[width=0.315\linewidth]{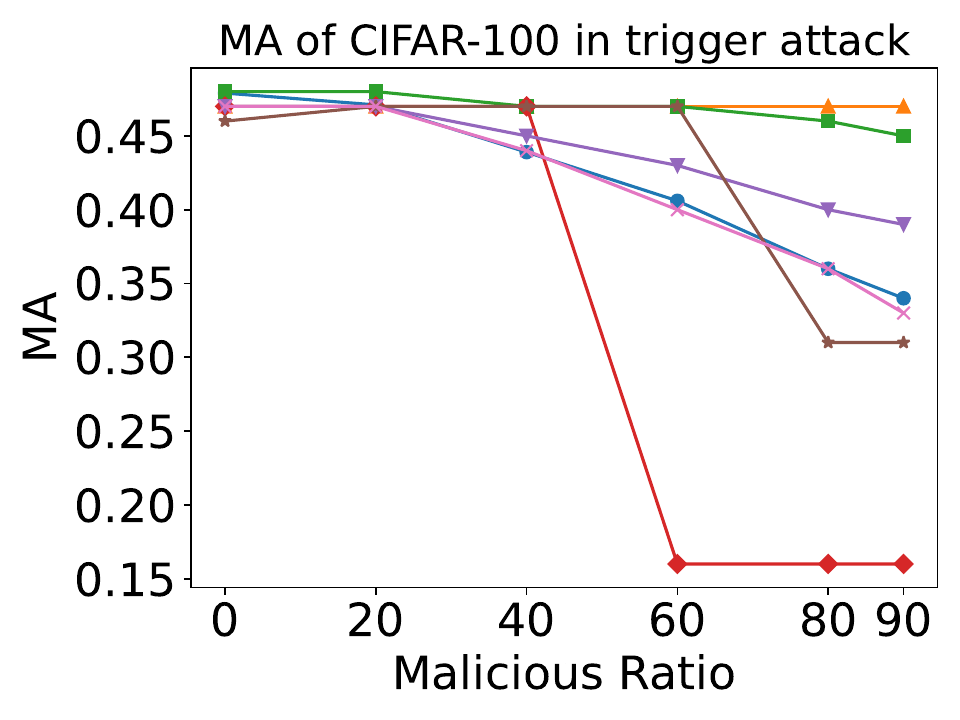}}
\end{center}}
\begin{center}
\includegraphics[width=1\linewidth]{figure/legend_fig3.pdf}
\end{center}
   \caption{MA of datasets in trigger attacks. \ref{fig:mamnist} shows how the MA of MNIST changes, \ref{fig:maCIFAR10} shows how MA of CIFAR-10 changes, and \ref{fig:maCIFAR100} shows how MA of CIFAR-100 changes.} 
\label{fig:maCIFAR101}
\end{figure*}

When there is no attack, as shown in Figure \ref{fig:mamnist} for the MNIST dataset, FLAME and NDC demonstrate good accuracy for the main task. FLAME utilizes clustering-based aggregation, which reduces noise via users update convergence as the proportion of malicious users increasing. NDC truncates update in surpassing a norm threshold, which mitigating impact from dissimilar user updates. This maintains the level of main-task accuracy (MA) closing to FedAvg. The limitation of Krum comes from its reliance on individual user models regards as the global model, and curtails data utilization. Our reduced main-task accuracy (MA) stems from the simplicity of the \texttt{Lenet} model. With two classifiers in the four-layer model, our method's division leads to a two-layer overall model.

As malicious user rate is increasing, FLTrust, NDC, FLAME, and FedAvg display slight decrease in MA but RSA sharply drops on the contrary. RSA emphasizes on updating direction to render it susceptible to malicious influences, which leads to pronounced MA reduction. The accuracy of Krum fluctuation stems from its user model selection, causing notable discrepancy in the final model for high malicious user proportions. Our method displays reduced sensitivity to malicious rate compared to RSA or Krum. Nonetheless, its effectiveness is limited on the MNIST dataset using the \texttt{Lenet} global model. This is elaborated in ``Limitation" Section.

For CIFAR-10 with \texttt{MobileNet} in Figure \ref{fig:maCIFAR10}, only RSA shows a lower level of effectiveness and the highest degree of decline among the backdoor defense strategies. It is because the strategy of RSA sends the direction of model updates as a parameter to the global model. If the global model is only aggregated by the update direction of the client model, it will greatly impact the MA. With the increasing proportion of malicious users, only FLTrust and FL-PLAS show a better MA, while others all have a certain degree of decline.

For the CIFAR-100 dataset with \texttt{ResNet-18} depicted in Figure \ref{fig:maCIFAR100}, when there is no attack, various defense strategies exhibit robust model aggregation without attacks except FLAME and Krum. With increasing malicious users, FL-PLAS, FLTrust, and RSA sustain MA within an acceptable range. FedAvg and NDC experience gradual MA decline, while FLAME and Krum face sharper drops at 40\% and 60\% respectively. This is attributed to FLAME and Krum selection or utilization of certain users' models for aggregation or as the global model. When the proportion of malicious users reaches a certain threshold, these methods favor malicious models, causing abrupt MA deterioration.

\begin{figure}[ht]
\begin{center}
\subfigcapskip=-3pt
\subfigure[semantic\label{fig:maseman}]{   \includegraphics[width=0.35\linewidth]{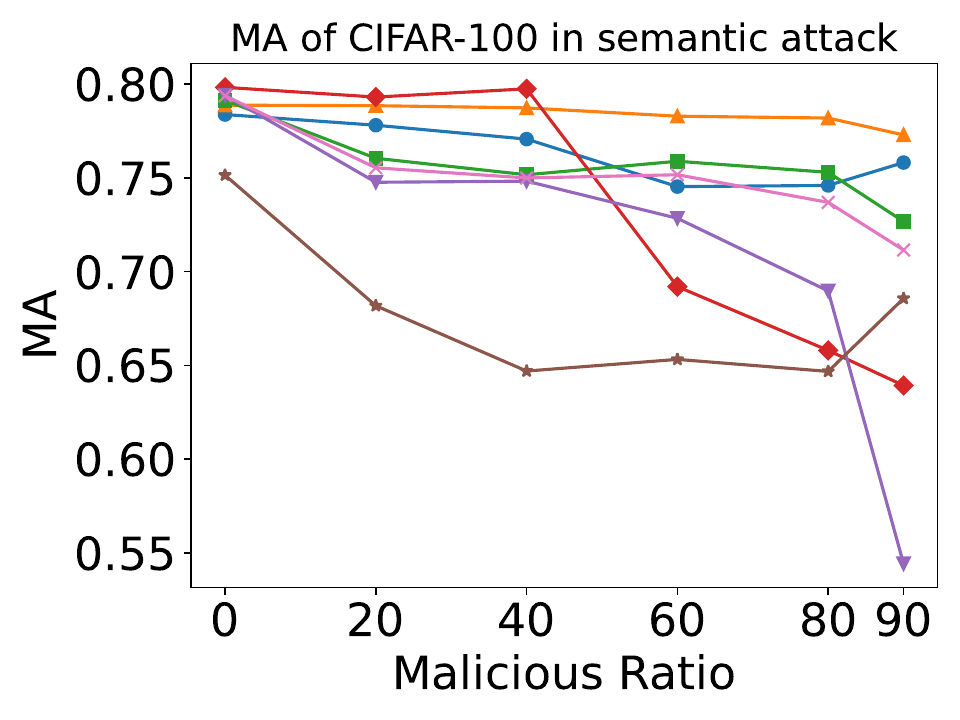}}
\hspace{20pt}
\subfigure[edge-case\label{fig:maedge}]{   \includegraphics[width=0.35\linewidth]{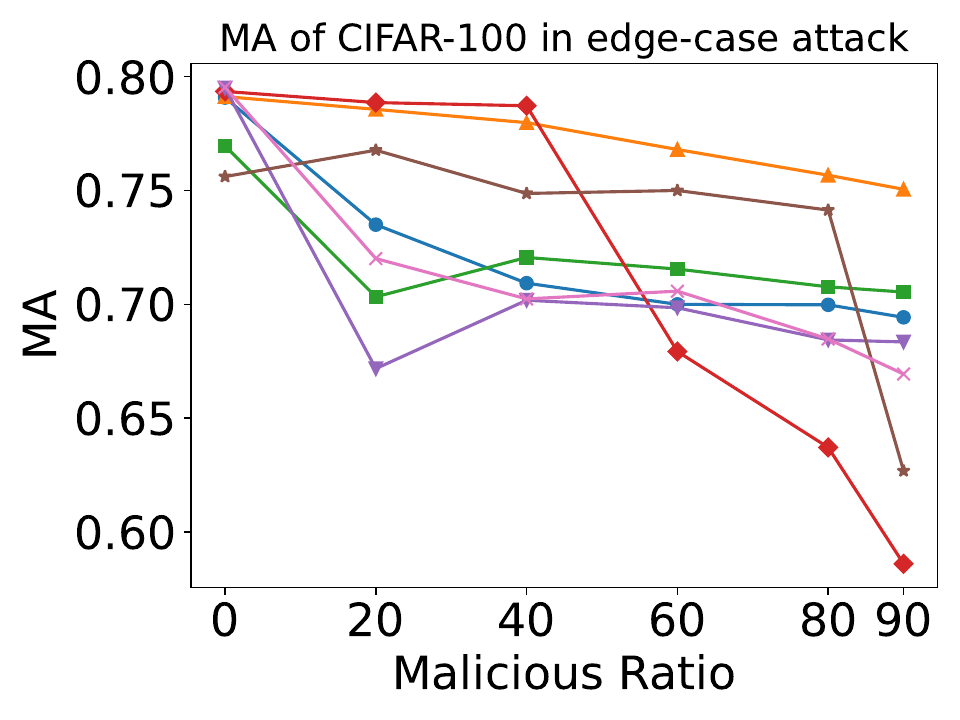}}
\end{center}
\begin{center}
\includegraphics[width=1\linewidth]{figure/legend_fig3.pdf}
\end{center}\
   \caption{MA of CIFAR-10 in two new types of attacks. \ref{fig:maseman} shows how the MA of CIFAR-10 changes in semantic attack, and \ref{fig:maedge} shows in edge-case attack.}
\label{fig:maedge0}
\end{figure}

In the semantic attack (Figure \ref{fig:maseman}), Krum's initial MA is lower due to it focuses on a subset model. With increasing malicious users, the drawbacks of FLAME and RSA gradually become more evident and the MA decreases rapidly. Meanwhile, the MA of FedAvg and FLTrust also decline in a certain degree, whereas our method FL-PLAS always performs the best MA.

In the edge-case attack shown in Figure \ref{fig:maedge}, only Krum has a lower MA when there is no attack. When the percentage of malicious users is less than 40\%, FLAME can still show a good MA, which means FLAME can detect malicious models. On the other hand, when the percentage of malicious users gradually increases, especially when the percentage of malicious users reaches 90\%, the MA of all these defense strategies is droping. In contrast, our method FL-PLAS defense method performs the best. Even when the proportion of malicious users is as high as 90\%, FL-PLAS still achieves a considerable reduction in MA, the results is shown in Table \ref{tab:0.9MA}.

\begin{table}[htb]
\footnotesize‌
\caption{The MA when there are 90\% malicious users.}
\vspace{-5mm}
\centering
\begin{center}
\scalebox{0.90}{
\subtable[Trigger attack]{
\setlength{\tabcolsep}{0.6mm}{
\begin{tabular}{ccccccccc}
\hline
&noattack&FedAvg&FLTrust&FLAME&RSA&Krum&NDC&FL-PLAS\\
\hline
MNIST     & 0.97     & \textbf{0.97} & 0.95          & \textbf{0.97} & 0.92 & 0.94 & \textbf{0.97} & 0.94          \\
CIFAR-10  & 0.79     & 0.65          & \textbf{0.79} & 0.72          & 0.44 & 0.68 & 0.66          & 0.76          \\
CIFAR-100 & 0.48     & 0.34          & 0.45          & 0.16          & 0.39 & 0.31 & 0.33          & \textbf{0.47} \\
\hline
\end{tabular}}}}
\scalebox{0.90}{
\subtable[New types of attack]{
\setlength{\tabcolsep}{0.6mm}{
\begin{tabular}{ccccccccc}
\hline
&noattack&FedAvg&FLTrust&FLAME&RSA&Krum&NDC&FL-PLAS\\
\hline
Semantic&0.78&0.76 &	0.73 &	0.64 &	0.54 &	0.69 &	0.71 &	\textbf{0.77}
\\
Edge-case&0.79&0.69&0.71&0.59&0.68&0.63&0.67&\textbf{0.75}\\
\hline
\end{tabular}}}}
\label{tab:0.9MA}
\end{center}
\end{table}

\subsubsection{Efficiency}
In contrast to the fundamental FedAvg, FL-PLAS doesn't necessitate additional computational resources or runtime space for users or servers. It entails partial user model aggregation and replacement of the server's global model by users for the aggregation's completed part. This process is resource-efficient, requiring no extra consumption. Conversely, most existing algorithms demand more computational power than FedAvg. Moreover, our method reduces server layers, minimizing data transfer volume and the risk of model compromise during transfers.

\begin{table}
\footnotesize‌
\caption{The screening complexity for seven aggregation methods. $\zeta$ denotes the number of local model update parameters, $M$ means the number of label classes and $\tau$ is the number of collected local model updates in each iteration.} 
\vspace{-1mm}
\begin{center}
\begin{tabular}{cc}
\hline
Method & Screen Complexity\\
\hline
\textbf{FedAvg} & $\textit{\textbf{O(0)}}$  \\
\textbf{FL-PLAS} &  $\textit{\textbf{O(0)}}$  \\
Multi-Krum & $O(\tau^2 \zeta)$  \\
Krum & $O(\tau^2 \zeta)$  \\
RFA & $O(\tau \zeta R^*)$  \\
RSA & $O(\tau \zeta)$  \\
NDC & $O(\tau \zeta)$  \\
\hline
\end{tabular}
\label{tab:complx}
\end{center}
\end{table}
In order to evaluate efficiency, we use screening complexity. It refers to how the server handles the received updates from local models in FL. Screening complexity involves screening, processing, and aggregating local model updates to ensure certain performance and efficiency. Thus we compare screening complexity of the proposed aggregation methods. In Table \ref{tab:complx}, the server using FL-PLAS doesn't need to screen the received local model updates, which is similar to FedAvg. Therefore, both of their screening complexities are $O(0)$. Krum and Multi-Krum compute the mutual distances between $\tau$ client local model updates. NDC and RSA prune and regularize $\tau$ local model updates respectively. RFA finds the geometric center by considering $\tau$ client local model updates until a defined condition is met. In summary, all the aforementioned methods classify malicious clients, hence they consider $\zeta$ parameter of local model updates. Overall, for evaluating screening complexity, as well as FedAvg, our FL-PLAS method performs better than others.


In Section \ref{evaluation}, conclusively, under a high malicious ratio, our method FL-PLAS has the best backdoor defense capability under traditional trigger attacks, semantic attacks, and edge-case attacks without requiring any user data on the server side. Similarly, compared with other defense methods, FL-PLAS also performing well in MNIST, CIFAR-10, and CIFAR-100, better than FedAvg, FLTrust, FLAME and even LFighter (see Appendix A), .etc. Also, we conducted experiments in Resnet-50 and Resnet-101 and the results were similar (see our code). In a word, it is good performance that FL-PLAS in using different network and defencing different adversarial methods.


\section{Discussion}
\label{discusssion}
\subsection{How Many Layers to Aggregate}

 In order to determine how many layers to aggregate in our partial layer aggregation method, we change the number of layers selected by our approach and analyze the results. We take the CIFAR-10 dataset using \texttt{MobileNet}, and the percentage of malicious users is 90\% as an example. We use $BA_{atk}=\frac{1}{MU}\sum{\frac{|B'_i|}{|B|}}$ to evaluate the depth of poisoning where $B'_i$ refers to the samples of the $i^{th}$ client whose labels are correctly predicted on the backdoor test set with the local model of malicious users and $||$ refers the size.

As we can see in Table \ref{tab:layer}, increasing the aggregation layer leads to higher main-task accuracy but also higher backdoor accuracy, indicating poorer defense. Therefore, the selection of aggregation layers needs to be analyzed for the corresponding network structures, rather than simply aggregating all feature extraction layers.

\begin{table}[htb]
\footnotesize‌
\caption{Effect of the number of aggregation layers. 
}
\vspace{-1mm}
\begin{center}
\setlength{\tabcolsep}{2.5mm}{
\scalebox{0.9}{
\begin{tabular}{ccccc}
\hline
Layers &MA&BA&BA$_{atk}$&BA$_{atk}$-BA\\
\hline
10&0.770&0.110&0.455&0.345\\
11&0.781&\textbf{0.109}&0.463&0.354\\
12&\textbf{0.785}&0.369&\textbf{0.769}&\textbf{0.400}\\
13 (classifier)&0.7897&0.900&0.900&-\\
\hline
\end{tabular}}}
\label{tab:layer}
\end{center}
\end{table}

\subsection{Convergence}

\begin{figure}[ht]
\begin{center}
\subfigcapskip=-3pt
\subfigure[Loss of MNIST\label{fig:con}]{\includegraphics[width=0.35\linewidth]{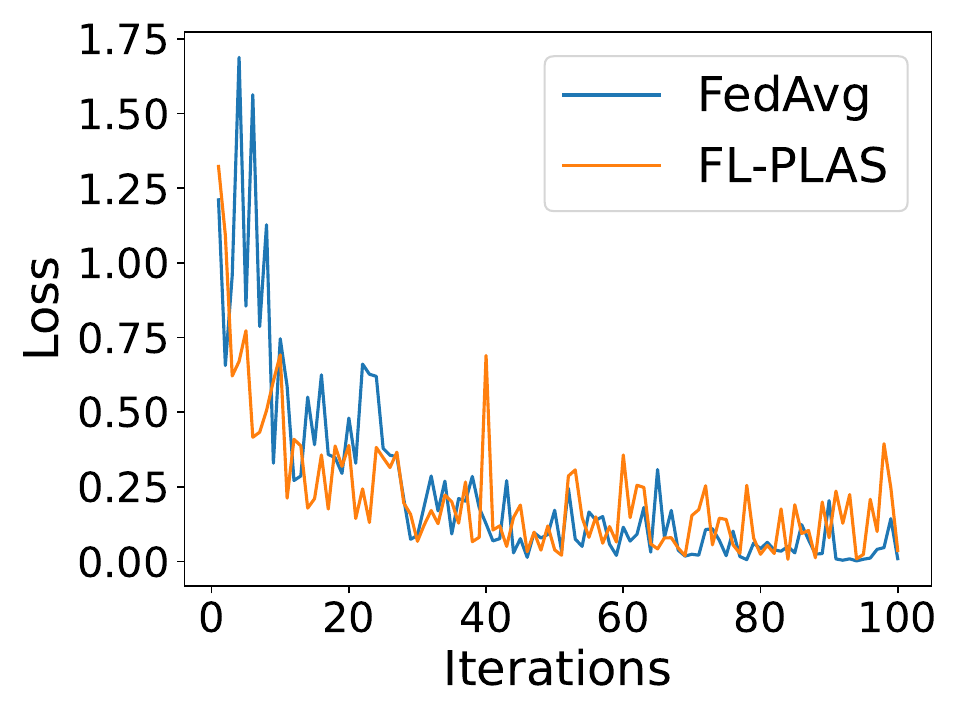}}
\hspace{25pt}
\subfigure[MA of CIFAR-10\label{fig:lim}]{\includegraphics[width=0.35\linewidth]{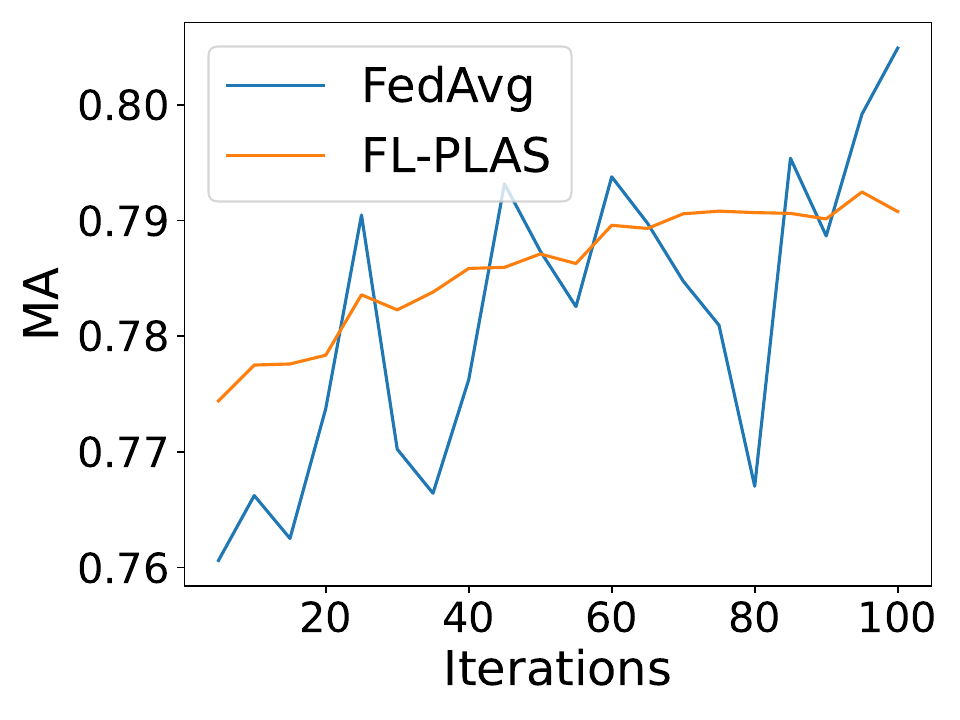}}
\end{center}
\caption{Effect of FL-PLAS on the model. \ref{fig:con} shows how the loss of MNIST changes in the trigger attack, and \ref{fig:lim} shows how MA if CIFAR-10 changes in the trigger attack in 100 iterations.}
\label{fig:dis}
\end{figure}

 In common network architectures like VGG9 (classifier scale:0.06\%), EfficientNet (0.09\%), GoogleNet (0.17\%), DenseNet (0.38\%),and SeNet (0.05\%), the parameters in the classifier impact on overall model is minimal. As the number of classifier layers is a small part of the model, not aggregating classifiers has a limited effect on convergence.  Figure \ref{fig:con} shows that the loss of FL-PLAS is 0.4 at the beginning, while the loss of FedAvg is 1.6. Their losses are both down to 0.1 after about 40 iterations.

Furthermore, Figure \ref{fig:con} illustrates FL-PLAS and FedAvg converging similarly in 40 iterations. In CIFAR-10 attack (Figure \ref{fig:lim}), convergence speeds match, but FedAvg fluctuates more, less robust than FL-PLAS.

If we use an over large classifier (i.e., split too many layers for classifier in the DNN), the model's convergence or accuracy would be impacted. Fortunately, we do not need to split too many layers for classifier to defend against backdoor attacks. It is because the backdoor activation neurons mainly aggregate on the last layers of DNN based on our experimental observation. So, we can keep the classifier small, which helps the model defend against backdoor attacks and maintain the model's accuracy.

\subsection{Limitation}
\label{limitation}

Based on experiments, FL-PLAS excels in robustness and main-task accuracy (MA) in most cases. However, some prior works like FLAME, NDC and FLTrust are better in certain cases while our method performs the best BA for robustness.

As for FLAME, when facing a high proportion of malicious users, the classification scheme may misclassify a moderate proportion of malicious users (more than 40\%) as normal users and consist them in model aggregation, even if their model accuracy on the MNIST dataset performs well as FedAvg. On the contrary, NDC relies on restricting model updates, resulting in higher MA but weaker defense effectiveness. As for FLTrust, although its MA is slightly better than our method on CIFAR-10, it performs slightly worse in BA. Moreover, the condition that server requires a certain amount of user data is quite controversial, whereas our method does not require any user data.

As for FL-PLAS, with a smaller model such as \texttt{LeNet}, the classifier occupies a higher proportion, and reduces the remaining part after clipping by our method. As the end-user's model learns less from others in the local classifier, using a lightweight neural network leads to decrease MA. It represents a trade-off between privacy and utility, while the BA and MA remain acceptable.

\subsection{Backdoor Attack and Byzantine Attack}

Our method focuses on the backdoor attack, while some prior works focus on the Byzantine attacks \cite{cao2020fltrust,blanchard2017machine,he2020byzantine,fang2020local,yin2018byzantine}. A Byzantine attack means an unknown number of malicious clients are omniscient, collude with each other, and send arbitrary vectors to the server to disrupt the learning process. A backdoor attack embeds hidden malicious behaviors into deep learning models, which only activate and cause misclassifications on inputs containing a specific trigger.

In federated learning, a Byzantine attack typically refers to an untargeted poisoning attack \cite{cao2020fltrust,dumford2020backdooring} that aims to destroy the accuracy of the global model. On the other hand, a backdoor attack aims to have the global model mislabel a specific portion of the samples without affecting the overall accuracy of the model. This type of attack is more threatening to the robustness and integrity of federated learning. Backdoor attacks can be carried out by a single attacker or multiple attackers, who may or may not collude with each other. If these attackers collude, the backdoor attack falls under the category of Byzantine attacks. Our method is designed to be insensitive to whether or not the malicious clients collude with each other. This means that our approach can defend against both Byzantine and non-Byzantine backdoor attacks.

\subsection{Personalized Federated Learning}
For federated learning, the non-IID characteristics of each user's local data will cause federated learning to face the problem of data heterogeneity. Federated learning itself needs to learn the information of participating users through distributed learning, which also causes the server to be unable to obtain the information of each user, which reduces the performance of the final model. A common solution to the problem of data heterogeneity is to personalize user data so that each device can obtain a higher-quality personalized model. Personalized Federated Learning (pFL) methods \cite{arivazhagan2019federated} not only strive to develop a global model but also aim to create a local personalized model for each client. They are more adaptable to the unique characteristics of each client's local dataset. As a result, pFL methods significantly outperform general FL methods in terms of prediction accuracy, particularly in practical scenarios with Non-IID data. pFL is primarily divided into full-model aggregation \cite{t2020personalized} and partial-model aggregation \cite{pillutla2022federated,tan2023pfedsim,qinpfl2023}. 

Although the concept of partial model sharing is similar with our method, it should be noted that our method is designed to target backdoor attacks. Unlike partial model aggregation, which may share parts of the model, layers, or parameters, our method targets specific layers. Additionally, there are significant differences in the design goals of these two approaches. The aim of pFL is to address the heterogeneity and individualization of data, focusing on localized personalized training. Conversely, our method lies in tackling the issue of a large proportion of malicious clients, preventing backdoor information from being propagated through aggregation.


\section{Conclusion} 
\label{conclusion} 

We propose and evaluate a new federated learning backdoor defense scheme called FL-PLAS. Based on some interesting insights, our method leverages the partial layer aggregation strategy to defend against backdoor attacks. We show that our method can handle cases where the percentage of malicious users is greater than 50\% without requiring additional user data. We evaluate the performance of our approach on three datasets under three backdoor attacks. The results demonstrate that our method can protect the local models of normal users from backdoor attacks, even when the percentage of malicious clients reaches 90\%, without any auxiliary dataset on the server. While our method may appear simple, the experimental results demonstrate the performance of FL-PLAS. It's the first time to propose partial layer aggregation for defending backdoors in federated-learning, and to address the challenge of training in large-presence of malicious clients. These conclusions suggest several important directions for future work, including expanding our research into natural language processing and audio processing, as well as exploring ways to mitigate model poisoning attacks.

\section*{Acknowledgment}
For the reproducibility of the proposed method, we have published our source code online at https://github.com/BESTICSP/FL-PLAS.

\section*{Appendix A}
\begin{figure*}[h]

\begin{center}
\subfigcapskip=-1pt
\subfigure[\label{fig:lfighterBM}]{\includegraphics[width=0.32\textwidth]{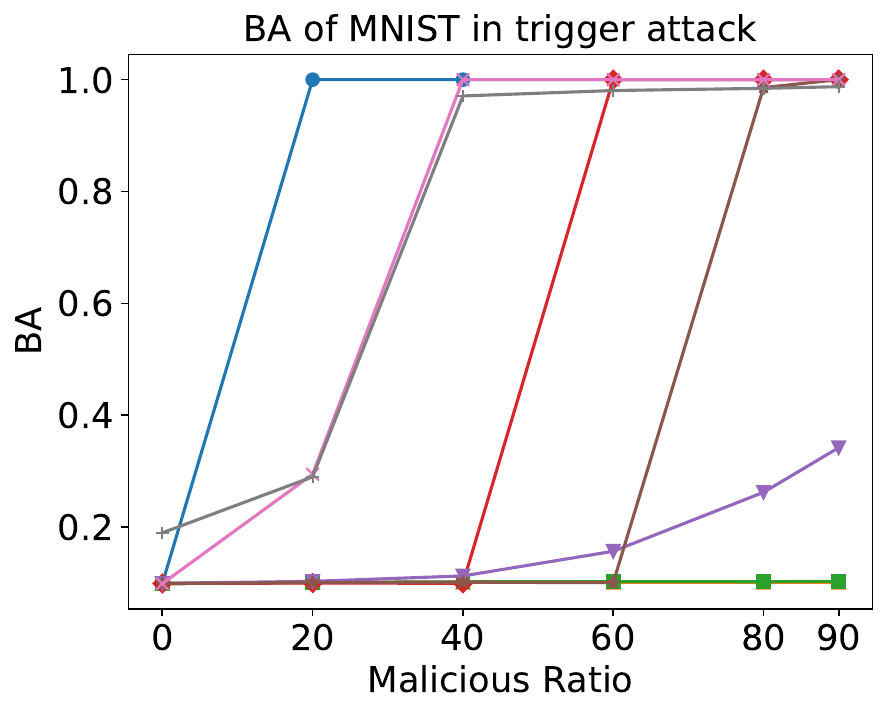}}
\hspace{20pt}
\subfigure[\label{fig:lfighterBC}]{\includegraphics[width=0.32\textwidth]{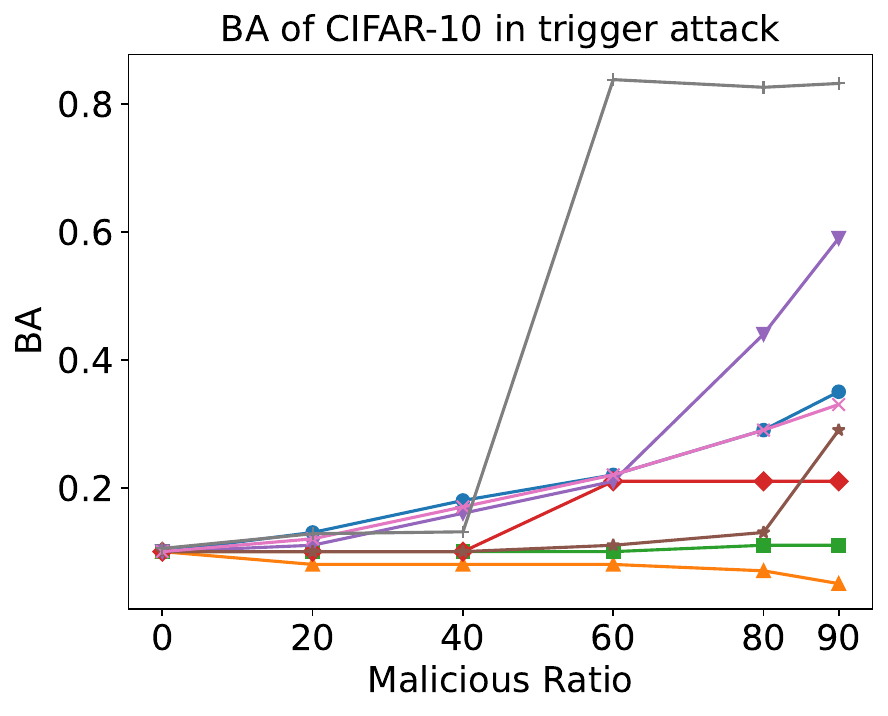}}
\subfigure[\label{fig:lfighterMM}]{\includegraphics[width=0.32\textwidth]{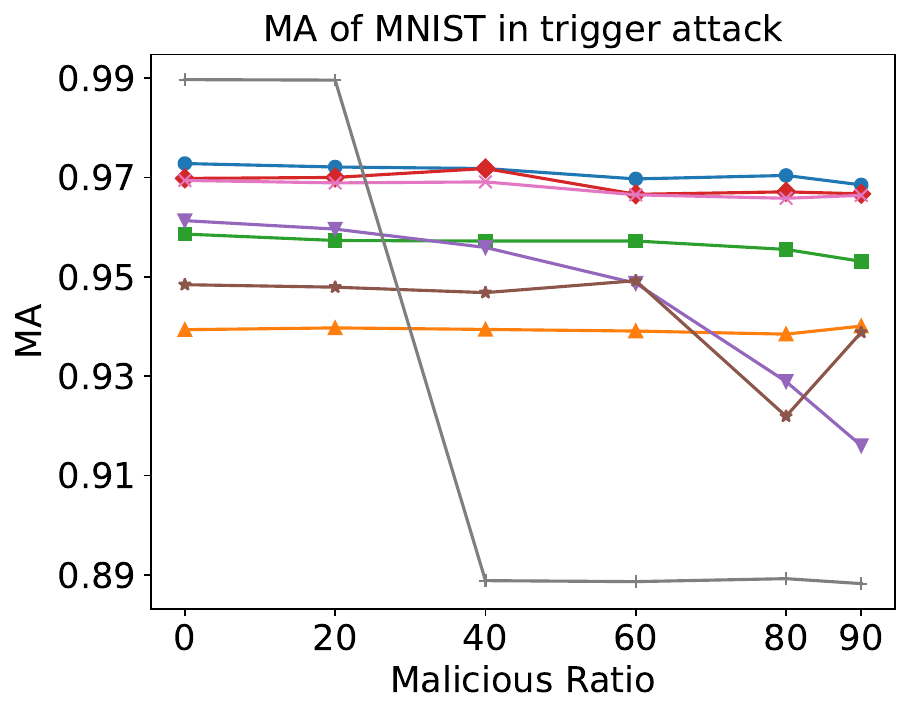}}
\hspace{20pt}
\subfigure[\label{fig:lfighterMC}]{\includegraphics[width=0.32\textwidth]{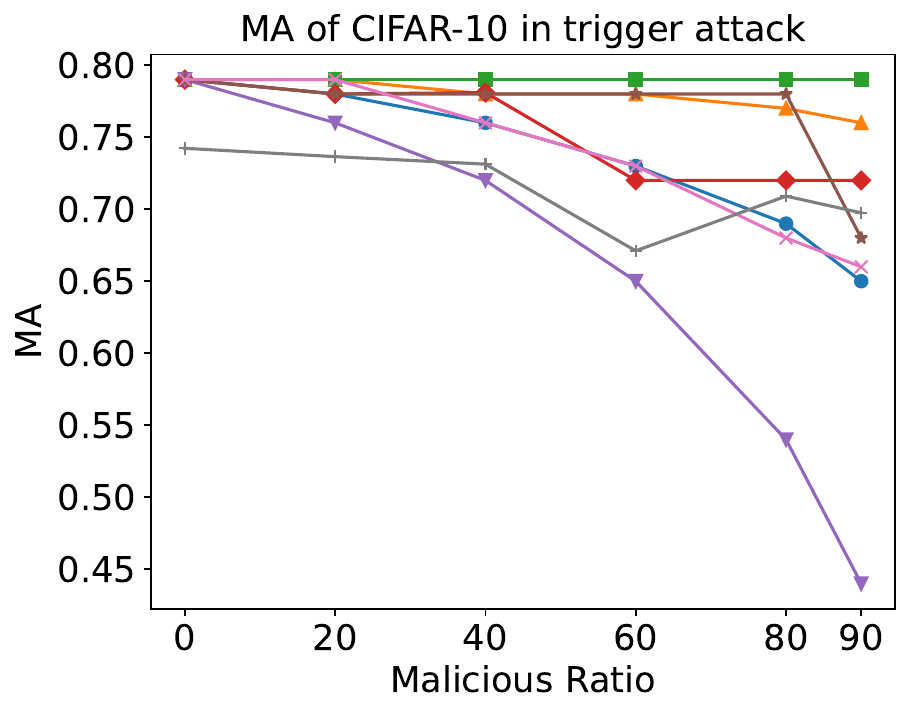}}
\end{center}
\begin{center}
\vspace{-5mm}
\includegraphics[width=1\linewidth]{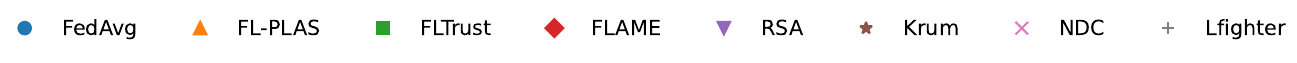}
\end{center}
\vspace{-5mm}
   \caption{BA and MA of datasets in trigger attacks. \ref{fig:lfighterBM} shows how the BA of malicious ratio changes under MNIST, \ref{fig:lfighterBC} shows how BA of CIFAR-10 changes, \ref{fig:lfighterMM} shows how MA of MNIST changes, and \ref{fig:lfighterMC} shows how MA of CIFAR-100 changes.} 
\end{figure*}

Figure \ref{fig:lfighterBM} and \ref{fig:lfighterBC} illustrate that when there are over 20\% and over 40\% malicious clients, respectively, LFighter exhibits a sudden increase in the backdoor attack (BA) rate on the MNIST and CIFAR10 datasets. This indicates that LFighter becomes ineffective in defending against backdoor attacks at these higher malicious client ratios. Within the 20\% and 40\% thresholds, its performance does not significantly differ from other defense methods. Most defense models can withstand attacks from a small number of malicious clients. However, at higher malicious client ratios, FL-PLAS consistently performs the best. On the other hand, Figure \ref{fig:lfighterMM} and \ref{fig:lfighterMC} show that LFighter maintains a high main task accuracy with a low percentage (below 20\%) of malicious clients. Yet, when the percentage exceeds 40\%, the model's performance rapidly deteriorates, and the main accuracy (MA) is significantly compromised by the backdoor data. Notably, with over 70\% malicious clients, LFighter's performance falls below that of the other defense methods.

\clearpage
\bibliographystyle{unsrt}  
\bibliography{references}

\begin{thebibliography}{10}

\bibitem{konevcny2016federated}
Jakub Kone{\v{c}}n{\`y}, H~Brendan McMahan, Felix~X Yu, Peter Richt{\'a}rik, Ananda~Theertha Suresh, and Dave Bacon.
\newblock Federated learning: Strategies for improving communication efficiency.
\newblock {\em arXiv preprint arXiv:1610.05492}, 2016.

\bibitem{mcmahan2017communication}
Brendan McMahan, Eider Moore, Daniel Ramage, Seth Hampson, and Blaise~Ag{\"{u}}era y~Arcas.
\newblock Communication-efficient learning of deep networks from decentralized data.
\newblock In {\em 20th International Conference on Artificial Intelligence and Statistics}. {PMLR}, 2017.

\bibitem{bouacida2021vulnerabilities}
Nader Bouacida and Prasant Mohapatra.
\newblock Vulnerabilities in federated learning.
\newblock {\em IEEE Access}, 9:63229--63249, 2021.

\bibitem{ozdayi2021defending}
Mustafa~Safa Ozdayi, Murat Kantarcioglu, and Yulia~R Gel.
\newblock Defending against backdoors in federated learning with robust learning rate.
\newblock In {\em Proceedings of the AAAI Conference on Artificial Intelligence}, volume~35, pages 9268--9276, 2021.

\bibitem{chen2017targeted}
Xinyun Chen, Chang Liu, Bo~Li, Kimberly Lu, and Dawn Song.
\newblock Targeted backdoor attacks on deep learning systems using data poisoning.
\newblock {\em arXiv preprint arXiv:1712.05526}, 2017.

\bibitem{gu2017badnets}
Tianyu Gu, Brendan Dolan-Gavitt, and Siddharth Garg.
\newblock Badnets: Identifying vulnerabilities in the machine learning model supply chain.
\newblock {\em arXiv preprint arXiv:1708.06733}, 2017.

\bibitem{bagdasaryan2020backdoor}
Eugene Bagdasaryan, Andreas Veit, Yiqing Hua, Deborah Estrin, and Vitaly Shmatikov.
\newblock How to backdoor federated learning.
\newblock In {\em 23rd International Conference on Artificial Intelligence and Statistics}, volume 108. {PMLR}, 2020.

\bibitem{tolpegin2020data}
Vale Tolpegin, Stacey Truex, Mehmet~Emre Gursoy, and Ling Liu.
\newblock Data poisoning attacks against federated learning systems.
\newblock In {\em Computer Security--ESORICS 2020: 25th European Symposium on Research in Computer Security, ESORICS 2020, Guildford, UK, September 14--18, 2020, Proceedings, Part I 25}, pages 480--501. Springer, 2020.

\bibitem{jebreel2024lfighter}
Najeeb~Moharram Jebreel, Josep Domingo-Ferrer, David S{\'a}nchez, and Alberto Blanco-Justicia.
\newblock Lfighter: Defending against the label-flipping attack in federated learning.
\newblock {\em Neural Networks}, 170:111--126, 2024.

\bibitem{wang2020attack}
Hongyi Wang, Kartik Sreenivasan, Shashank Rajput, Harit Vishwakarma, Saurabh Agarwal, Jy-yong Sohn, Kangwook Lee, and Dimitris Papailiopoulos.
\newblock Attack of the tails: Yes, you really can backdoor federated learning.
\newblock {\em arXiv preprint arXiv:2007.05084}, 2020.

\bibitem{fang2023vulnerability}
Pei Fang and Jinghui Chen.
\newblock On the vulnerability of backdoor defenses for federated learning.
\newblock In {\em AAAI Conference on Artificial Intelligence}, 2023.

\bibitem{zhuang2024backdoor}
Haomin Zhuang, Mingxian Yu, Hao Wang, Yang Hua, Jian Li, and Xu~Yuan.
\newblock Backdoor federated learning by poisoning backdoor-critical layers.
\newblock 2024.

\bibitem{lyu2020threats}
Lingjuan Lyu, Han Yu, and Qiang Yang.
\newblock Threats to federated learning: A survey.
\newblock {\em arXiv preprint arXiv:2003.02133}, 2020.

\bibitem{li2019rsa}
Liping Li, Wei Xu, Tianyi Chen, Georgios~B. Giannakis, and Qing Ling.
\newblock {RSA:} byzantine-robust stochastic aggregation methods for distributed learning from heterogeneous datasets.
\newblock In {\em 33rd {AAAI} Conference on Artificial Intelligence}, 2019.

\bibitem{sun2019can}
Ziteng Sun, Peter Kairouz, Ananda~Theertha Suresh, and H~Brendan McMahan.
\newblock Can you really backdoor federated learning?
\newblock {\em arXiv preprint arXiv:1911.07963}, 2019.

\bibitem{li2020learning}
Suyi Li, Yong Cheng, Wei Wang, Yang Liu, and Tianjian Chen.
\newblock Learning to detect malicious clients for robust federated learning.
\newblock {\em arXiv preprint arXiv:2002.00211}, 2020.

\bibitem{cao2020fltrust}
Xiaoyu Cao, Minghong Fang, Jia Liu, and Neil~Zhenqiang Gong.
\newblock Fltrust: Byzantine-robust federated learning via trust bootstrapping.
\newblock In {\em Network and Distributed Systems Security (NDSS) Symposium}, 2021.

\bibitem{nguyen2021flame}
Thien~Duc Nguyen, Phillip Rieger, Roberta De~Viti, Huili Chen, Bj{\"o}rn~B Brandenburg, Hossein Yalame, Helen M{\"o}llering, Hossein Fereidooni, Samuel Marchal, Markus Miettinen, et~al.
\newblock $\{$FLAME$\}$: Taming backdoors in federated learning.
\newblock In {\em 31st USENIX Security Symposium (USENIX Security 22)}, pages 1415--1432, 2022.

\bibitem{blanchard2017machine}
Peva Blanchard, El~Mahdi~El Mhamdi, Rachid Guerraoui, and Julien Stainer.
\newblock Machine learning with adversaries: Byzantine tolerant gradient descent.
\newblock In {\em 30th Annual Conference on Neural Information Processing Systems}, 2017.

\bibitem{dumford2020backdooring}
Jacob Dumford and Walter Scheirer.
\newblock Backdooring convolutional neural networks via targeted weight perturbations.
\newblock In {\em 2020 IEEE International Joint Conference on Biometrics (IJCB)}. IEEE, 2020.

\bibitem{fang2021data}
Minghong Fang, Minghao Sun, Qi~Li, Neil Gong, Jin Tian, and Jia Liu.
\newblock Data poisoning attacks and defenses to crowdsourcing systems.
\newblock In {\em Proceedings of the Web Conference}, pages 969--980, 2021.

\bibitem{fang2018poisoning}
Minghong Fang, Guolei Yang, Neil~Zhenqiang Gong, and Jia Liu.
\newblock Poisoning attacks to graph-based recommender systems.
\newblock In {\em Proceedings of the 34th Annual Computer Security Applications Conference}, pages 381--392, 2018.

\bibitem{yang2017fake}
Guolei Yang, Neil~Zhenqiang Gong, and Ying Cai.
\newblock Fake co-visitation injection attacks to recommender systems.
\newblock In {\em NDSS}, 2017.

\bibitem{nelson2008exploiting}
Blaine Nelson, Marco Barreno, Fuching~Jack Chi, Anthony~D Joseph, Benjamin~IP Rubinstein, Udam Saini, Charles~A Sutton, J~Doug Tygar, and Kai Xia.
\newblock Exploiting machine learning to subvert your spam filter.
\newblock {\em LEET}, 8:1--9, 2008.

\bibitem{kairouz2019advances}
Peter Kairouz, H~Brendan McMahan, Brendan Avent, Aur{\'e}lien Bellet, Mehdi Bennis, Arjun~Nitin Bhagoji, Keith Bonawitz, Zachary Charles, Graham Cormode, Rachel Cummings, et~al.
\newblock Advances and open problems in federated learning.
\newblock {\em arXiv preprint arXiv:1912.04977}, 2019.

\bibitem{li2020backdoor}
Yiming Li, Baoyuan Wu, Yong Jiang, Zhifeng Li, and Shu-Tao Xia.
\newblock Backdoor learning: A survey.
\newblock {\em arXiv preprint arXiv:2007.08745}, 2020.

\bibitem{he2020byzantine}
Lie He, Sai~Praneeth Karimireddy, and Martin Jaggi.
\newblock Byzantine-robust learning on heterogeneous datasets via resampling.
\newblock {\em arXiv preprint arXiv:2006.09365}, 2020.

\bibitem{hsu2019measuring}
Tzu-Ming~Harry Hsu, Hang Qi, and Matthew Brown.
\newblock Measuring the effects of non-identical data distribution for federated visual classification.
\newblock {\em arXiv preprint arXiv:1909.06335}, 2019.

\bibitem{goodfellow2014explaining}
Ian~J Goodfellow, Jonathon Shlens, and Christian Szegedy.
\newblock Explaining and harnessing adversarial examples.
\newblock {\em arXiv preprint arXiv:1412.6572}, 2014.

\bibitem{papernot2016limitations}
Nicolas Papernot, Patrick McDaniel, Somesh Jha, Matt Fredrikson, Z~Berkay Celik, and Ananthram Swami.
\newblock The limitations of deep learning in adversarial settings.
\newblock In {\em 2016 IEEE European symposium on security and privacy}. IEEE, 2016.

\bibitem{hong2019terminal}
Sanghyun Hong, Pietro Frigo, Yi{\u{g}}itcan Kaya, Cristiano Giuffrida, and Tudor Dumitraș.
\newblock Terminal brain damage: Exposing the graceless degradation in deep neural networks under hardware fault attacks.
\newblock In {\em 28th USENIX Security Symposium}, 2019.

\bibitem{rubinstein2009antidote}
Benjamin~IP Rubinstein, Blaine Nelson, Ling Huang, Anthony~D Joseph, Shing-hon Lau, Satish Rao, Nina Taft, and J~Doug Tygar.
\newblock Antidote: understanding and defending against poisoning of anomaly detectors.
\newblock In {\em 9th ACM SIGCOMM Conference on Internet Measurement}, 2009.

\bibitem{shafahi2018poison}
Ali Shafahi, W.~Ronny Huang, Mahyar Najibi, Octavian Suciu, Christoph Studer, Tudor Dumitras, and Tom Goldstein.
\newblock Poison frogs! targeted clean-label poisoning attacks on neural networks.
\newblock In {\em 31st Annual Conference on Neural Information Processing Systems, NeurIPS}, pages 6106--6116, 2018.

\bibitem{suciu2018does}
Octavian Suciu, Radu Marginean, Yigitcan Kaya, Hal Daume~III, and Tudor Dumitras.
\newblock When does machine learning fail? generalized transferability for evasion and poisoning attacks.
\newblock In {\em 27th USENIX Security Symposium}, 2018.

\bibitem{wang2019attacking}
Binghui Wang and Neil~Zhenqiang Gong.
\newblock Attacking graph-based classification via manipulating the graph structure.
\newblock In {\em ACM CCS}, 2019.

\bibitem{biggio2012poisoning}
Battista Biggio, Blaine Nelson, and Pavel Laskov.
\newblock Poisoning attacks against support vector machines.
\newblock In {\em Proceedings of the 29th International Conference on Machine Learning, {ICML} 2012, Edinburgh, Scotland, UK, June 26 - July 1, 2012}. icml.cc / Omnipress, 2012.

\bibitem{jagielski2018manipulating}
Matthew Jagielski, Alina Oprea, Battista Biggio, Chang Liu, Cristina Nita-Rotaru, and Bo~Li.
\newblock Manipulating machine learning: Poisoning attacks and countermeasures for regression learning.
\newblock In {\em 2018 IEEE Symposium on Security and Privacy (SP)}, pages 19--35. IEEE, 2018.

\bibitem{li2016data}
Bo~Li, Yining Wang, Aarti Singh, and Yevgeniy Vorobeychik.
\newblock Data poisoning attacks on factorization-based collaborative filtering.
\newblock In Daniel~D. Lee, Masashi Sugiyama, Ulrike von Luxburg, Isabelle Guyon, and Roman Garnett, editors, {\em Advances in Neural Information Processing Systems 29: Annual Conference on Neural Information Processing Systems 2016, December 5-10, 2016, Barcelona, Spain}, pages 1885--1893, 2016.

\bibitem{munoz2017towards}
Luis Mu{\~n}oz-Gonz{\'a}lez, Battista Biggio, Ambra Demontis, Andrea Paudice, Vasin Wongrassamee, Emil~C Lupu, and Fabio Roli.
\newblock Towards poisoning of deep learning algorithms with back-gradient optimization.
\newblock In {\em Proceedings of the 10th ACM Workshop on Artificial Intelligence and Security}, pages 27--38, 2017.

\bibitem{xiao2015feature}
Huang Xiao, Battista Biggio, Gavin Brown, Giorgio Fumera, Claudia Eckert, and Fabio Roli.
\newblock Is feature selection secure against training data poisoning?
\newblock In {\em Proceedings of the 32nd International Conference on Machine Learning, {ICML}}, volume~37, pages 1689--1698. JMLR.org, 2015.

\bibitem{fang2020local}
Minghong Fang, Xiaoyu Cao, Jinyuan Jia, and Neil Gong.
\newblock Local model poisoning attacks to byzantine-robust federated learning.
\newblock In {\em 29th $\{$USENIX$\}$ Security Symposium}, 2020.

\bibitem{baruch2019little}
Gilad Baruch, Moran Baruch, and Yoav Goldberg.
\newblock A little is enough: Circumventing defenses for distributed learning.
\newblock In Hanna~M. Wallach, Hugo Larochelle, Alina Beygelzimer, Florence d'Alch{\'{e}}{-}Buc, Emily~B. Fox, and Roman Garnett, editors, {\em Advances in Neural Information Processing Systems 32: Annual Conference on Neural Information Processing Systems 2019, NeurIPS 2019, December 8-14, 2019, Vancouver, BC, Canada}, pages 8632--8642, 2019.

\bibitem{xie2020fall}
Cong Xie, Oluwasanmi Koyejo, and Indranil Gupta.
\newblock Fall of empires: Breaking byzantine-tolerant {SGD} by inner product manipulation.
\newblock In {\em Proceedings of the Thirty-Fifth Conference on Uncertainty in Artificial Intelligence, {UAI}}, volume 115, pages 261--270. {AUAI} Press, 2019.

\bibitem{bhagoji2019analyzing}
Arjun~Nitin Bhagoji, Supriyo Chakraborty, Prateek Mittal, and Seraphin~B. Calo.
\newblock Analyzing federated learning through an adversarial lens.
\newblock In Kamalika Chaudhuri and Ruslan Salakhutdinov, editors, {\em Proceedings of the 36th International Conference on Machine Learning, {ICML} 2019, 9-15 June 2019, Long Beach, California, {USA}}, volume~97 of {\em Proceedings of Machine Learning Research}, pages 634--643. {PMLR}, 2019.

\bibitem{xie2019dba}
Chulin Xie, Keli Huang, Pin{-}Yu Chen, and Bo~Li.
\newblock {DBA:} distributed backdoor attacks against federated learning.
\newblock In {\em 8th International Conference on Learning Representations, {ICLR} 2020, Addis Ababa, Ethiopia, April 26-30, 2020}. OpenReview.net, 2020.

\bibitem{liu2017neural}
Yuntao Liu, Yang Xie, and Ankur Srivastava.
\newblock Neural trojans.
\newblock In {\em International Conference on Computer Design}. IEEE, 2017.

\bibitem{doan2020februus}
Bao~Gia Doan, Ehsan Abbasnejad, and Damith~C Ranasinghe.
\newblock Februus: Input purification defense against trojan attacks on deep neural network systems.
\newblock In {\em Annual Computer Security Applications Conference}, 2020.

\bibitem{zhao2020bridging}
Pu~Zhao, Pin-Yu Chen, Payel Das, Karthikeyan~Natesan Ramamurthy, and Xue Lin.
\newblock Bridging mode connectivity in loss landscapes and adversarial robustness.
\newblock {\em arXiv preprint arXiv:2005.00060}, 2020.

\bibitem{liu2018fine}
Kang Liu, Brendan Dolan-Gavitt, and Siddharth Garg.
\newblock Fine-pruning: Defending against backdooring attacks on deep neural networks.
\newblock In {\em International Symposium on Research in Attacks, Intrusions, and Defenses}. Springer, 2018.

\bibitem{tran2018spectral}
Brandon Tran, Jerry Li, and Aleksander Madry.
\newblock Spectral signatures in backdoor attacks.
\newblock {\em Advances in neural information processing systems}, 31, 2018.

\bibitem{chen2018detecting}
Bryant Chen, Wilka Carvalho, Nathalie Baracaldo, Heiko Ludwig, Benjamin Edwards, Taesung Lee, Ian Molloy, and Biplav Srivastava.
\newblock Detecting backdoor attacks on deep neural networks by activation clustering.
\newblock {\em arXiv preprint arXiv:1811.03728}, 2018.

\bibitem{tang2021demon}
Di~Tang, XiaoFeng Wang, Haixu Tang, and Kehuan Zhang.
\newblock Demon in the variant: Statistical analysis of dnns for robust backdoor contamination detection.
\newblock In {\em 30th USENIX Security Symposium}, 2021.

\bibitem{andreina2021baffle}
Sebastien Andreina, Giorgia~Azzurra Marson, Helen M{\"o}llering, and Ghassan Karame.
\newblock Baffle: Backdoor detection via feedback-based federated learning.
\newblock In {\em 41st International Conference on Distributed Computing Systems}, pages 852--863. IEEE, 2021.

\bibitem{gao2019strip}
Yansong Gao, Change Xu, Derui Wang, Shiping Chen, Damith~C Ranasinghe, and Surya Nepal.
\newblock Strip: A defence against trojan attacks on deep neural networks.
\newblock In {\em Proceedings of the 35th Annual Computer Security Applications Conference}, 2019.

\bibitem{subedar2019deep}
Mahesh Subedar, Nilesh Ahuja, Ranganath Krishnan, Ibrahima~J Ndiour, and Omesh Tickoo.
\newblock Deep probabilistic models to detect data poisoning attacks.
\newblock {\em arXiv preprint arXiv:1912.01206}, 2019.

\bibitem{jin2020unified}
Kaidi Jin, Tianwei Zhang, Chao Shen, Yufei Chen, Ming Fan, Chenhao Lin, and Ting Liu.
\newblock A unified framework for analyzing and detecting malicious examples of dnn models.
\newblock {\em arXiv preprint arXiv:2006.14871}, 2020.

\bibitem{zhao2022fedinv}
Bo~Zhao, Peng Sun, Tao Wang, and Keyu Jiang.
\newblock Fedinv: Byzantine-robust federated learning by inversing local model updates.
\newblock In {\em 36th AAAI Conference on Artificial Intelligence}, 2022.

\bibitem{zhang2023flip}
Kaiyuan Zhang, Guanhong Tao, Qiuling Xu, Siyuan Cheng, Shengwei An, Yingqi Liu, Shiwei Feng, Guangyu Shen, Pin-Yu Chen, Shiqing Ma, et~al.
\newblock Flip: A provable defense framework for backdoor mitigation in federated learning.
\newblock In {\em International Conference on Learning Representations (ICLR)}, 2023.

\bibitem{raza2022using}
Ali Raza, Shujun Li, Kim-Phuc Tran, and Ludovic Koehl.
\newblock Using anomaly detection to detect poisoning attacks in federated learning applications.
\newblock {\em arXiv preprint arXiv:2207.08486}, 2022.

\bibitem{collins2021exploiting}
Liam Collins, Hamed Hassani, Aryan Mokhtari, and Sanjay Shakkottai.
\newblock Exploiting shared representations for personalized federated learning.
\newblock In {\em International conference on machine learning}, pages 2089--2099. PMLR, 2021.

\bibitem{chen2024efficient}
Zhixiong Chen, Wenqiang Yi, Hyundong Shin, Arumugam Nallanathan, and Geoffrey~Ye Li.
\newblock Efficient wireless federated learning with partial model aggregation.
\newblock {\em IEEE Transactions on Communications}, 2024.

\bibitem{pillutla2022federated}
Krishna Pillutla, Kshitiz Malik, Abdel-Rahman Mohamed, Mike Rabbat, Maziar Sanjabi, and Lin Xiao.
\newblock Federated learning with partial model personalization.
\newblock In {\em International Conference on Machine Learning}, pages 17716--17758. PMLR, 2022.

\bibitem{arivazhagan2019federated}
Manoj~Ghuhan Arivazhagan, Vinay Aggarwal, Aaditya~Kumar Singh, and Sunav Choudhary.
\newblock Federated learning with personalization layers.
\newblock {\em arXiv preprint arXiv:1912.00818}, 2019.

\bibitem{t2020personalized}
Canh T~Dinh, Nguyen Tran, and Josh Nguyen.
\newblock Personalized federated learning with moreau envelopes.
\newblock {\em Advances in Neural Information Processing Systems}, 33:21394--21405, 2020.

\bibitem{tan2023pfedsim}
Jiahao Tan, Yipeng Zhou, Gang Liu, Jessie~Hui Wang, and Shui Yu.
\newblock \textbf{pFedSim}: Similarity-aware model aggregation towards personalized federated learning.
\newblock {\em arXiv preprint arXiv:2305.15706}, 2023.

\bibitem{gao2020end}
Yansong Gao, Minki Kim, Sharif Abuadbba, Yeonjae Kim, Chandra Thapa, Kyuyeon Kim, Seyit~A Camtepe, Hyoungshick Kim, and Surya Nepal.
\newblock End-to-end evaluation of federated learning and split learning for internet of things.
\newblock {\em arXiv preprint arXiv:2003.13376}, 2020.

\bibitem{qinpfl2023}
Zeyu Qin, Liuyi Yao, Daoyuan Chen, Yaliang Li, Bolin Ding, and Minhao Cheng.
\newblock Revisiting personalized federated learning: Robustness against backdoor attacks.
\newblock In {\em Proceedings of the 29th ACM SIGKDD Conference on Knowledge Discovery and Data Mining}, KDD '23, page 4743–4755, New York, NY, USA, 2023. Association for Computing Machinery.

\bibitem{huang2020one}
Shanjiaoyang Huang, Weiqi Peng, Zhiwei Jia, and Zhuowen Tu.
\newblock One-pixel signature: Characterizing cnn models for backdoor detection.
\newblock In {\em European Conference on Computer Vision}, pages 326--341. Springer, 2020.

\bibitem{xmam}
Jianyi Zhang, Fangjiao Zhang, Qichao Jin, Zhiqiang Wang, Xiaodong Lin, and Xiali Hei.
\newblock Xmam: X-raying models with a matrix to reveal backdoor attacks for federated learning.
\newblock {\em Digital Communications and Networks}, 10(4):1154--1167, 2024.

\bibitem{lecun1998gradient}
Yann LeCun, L{\'e}on Bottou, Yoshua Bengio, and Patrick Haffner.
\newblock Gradient-based learning applied to document recognition.
\newblock {\em Proceedings of the IEEE}, 86(11):2278--2324, 1998.

\bibitem{lecun1989handwritten}
Yann LeCun, Bernhard Boser, John Denker, Donnie Henderson, Richard Howard, Wayne Hubbard, and Lawrence Jackel.
\newblock Handwritten digit recognition with a back-propagation network.
\newblock {\em Advances in neural information processing systems}, 2, 1989.

\bibitem{krizhevsky2009learning}
Alex Krizhevsky, Geoffrey Hinton, et~al.
\newblock Learning multiple layers of features from tiny images.
\newblock 2009.

\bibitem{he2016deep}
Kaiming He, Xiangyu Zhang, Shaoqing Ren, and Jian Sun.
\newblock Deep residual learning for image recognition.
\newblock In {\em IEEE conference on computer vision and pattern recognition}, 2016.

\bibitem{howard2017mobilenets}
Andrew~G Howard, Menglong Zhu, Bo~Chen, Dmitry Kalenichenko, Weijun Wang, Tobias Weyand, Marco Andreetto, and Hartwig Adam.
\newblock Mobilenets: Efficient convolutional neural networks for mobile vision applications.
\newblock {\em arXiv preprint arXiv:1704.04861}, 2017.

\bibitem{yin2018byzantine}
Dong Yin, Yudong Chen, Ramchandran Kannan, and Peter Bartlett.
\newblock Byzantine-robust distributed learning: Towards optimal statistical rates.
\newblock In {\em ICML}, pages 5650--5659. PMLR, 2018.

\end{thebibliography}

\end{document}